\documentstyle[preprint,aps,epsfig]{revtex}
\def\al{\alpha}

\def\sig{\sigma}

\def\bold#1{{\bf #1}}

\def\eps{\epsilon}

\def\vpi{\varpi}

\def\txs#1{\textstyle #1}
\def\lan{<\!}
\def\ran{\!>}
\def\beq{\begin{equation}}
\def\eeq{\end{equation}}
\def\bea{\begin{eqnarray}}
\def\eea{\end{eqnarray}}
\def\nn{\nonumber}
\def\eps{\epsilon}

\newcommand{\bk}{\bbox{k}}
\newcommand{\bp}{\bbox{p}}
\newcommand{\bq}{\bbox{q}}

\newcommand{\bP}{\bbox{P}}

\newcommand{\br}{\bbox{r}}

\newcommand{\bsigma}{\bbox{\sigma}}
\newcommand{\bnabla}{\bbox{\nabla}}

\newcommand{\btau}{\bbox{\tau}}
\newcommand{\bcdot}{\bbox{\cdot}}
\def\aq{\bar{q}}

\def\e{{\cal E}}

\def\caln{{\cal N}}
\def\calo{{\cal O}}
\def\sig{\sigma}
\def\ft{\rm FT}

\def\aq{\bar{q}}
\def\al{\alpha}
\def\qq{\rm qq}

\def\bm#1{\mbox{\boldmath{$#1$}}}
\begin{document}
\begin{center}
{\large \bf Mapping of composite hadrons into elementary hadrons}\\
{\large \bf and effective hadronic Hamiltonians}
\end{center}
\vskip 5.0mm
\begin{center}
D. HADJIMICHEF$^{\,a}$, G. KREIN$^{\,*b}$, S. SZPIGEL$^{\,c}$ AND
J.S. DA VEIGA$^{\, d}$
\end{center}

\centerline{$^a${\it Dep. de F\'{\i}sica, Funda\c{c}\~ao Universidade do 
Rio Grande, 96201-900 Rio Grande-RS, Brazil }      }
\centerline{$^b$ {\it Institut f\"{u}r Kernphysik, Universit\"{a}t Mainz,
D-55099 Mainz, Germany}}
\centerline{$^c${\it Department of Physics, The Ohio State University,
Columbus, OH 43210, USA}}
\centerline{$^d${\it Instituto de F\'{\i}sica, Universidade de S\~ao Paulo,
01452-900 S\~ao Paulo-SP, Brazil} }
\begin{abstract}
A mapping technique is used to derive in the context of constituent quark 
models effective Hamiltonians that involve explicit hadron degrees of freedom.
The technique is based on the ideas of mapping between physical and ideal
Fock spaces and shares similarities with the quasiparticle method of Weinberg. 
Starting with the Fock-space representation of single-hadron states, a change 
of representation is implemented by a unitary transformation such that 
composites are redescribed by elementary Bose and Fermi field operators in an 
extended Fock space. When the unitary transformation is applied to the 
microscopic quark Hamiltonian, effective, hermitian Hamiltonians with a clear 
physical interpretation are obtained. Applications and comparisons with other 
composite-particle formalisms of the recent literature are made using the 
nonrelativistic quark model. 
\end{abstract}

\vspace{0.5cm}

\vspace{1.0cm}
\noindent
$^{\ast}$
Alexander von Humboldt Research Fellow\\
Permanent address: Instituto de F\'{\i}sica Te\'orica, Universidade Estadual 
Paulista\\
Rua Pamplona 145, 01405-900 S\~ao Paulo-SP, Brazil

%
\newpage 
%
%
%
%
%
%
%
%
%
%
%
\section{Introduction}
 
A great variety of mapping techniques are presented in the literature. In
nuclear physics they are used to treat collective oscillations of nuclei. 
Although  available since a long time, there have been attempts only very 
recently to extend these techniques to hadronic physics, in particular, to
constituent quark models. The pioneering work originates from Zhu et
al.~\cite{Zhu} and Pittel et al.~\cite{PittEngDukRi}. Zhu et al. use the
Composite Particle Representation (CPR) developed by Wu, Feng and 
collaborators~\cite{WuFeng} in the context of nuclear physics, for studying
the baryon spectrum in the nonrelativistic quark model. Pittel et 
al. used the Dyson mapping~\cite{dys} to obtain an 
effective hadron-hadron interaction from a schematic quark model. A continued
effort following this is contained in 
Refs.~\cite{{DukPitt},{PittAriDukFra},{GRStoPittDuk},{StoPittDuk}}. Related 
work is contained in 
Refs.~\cite{{Nadj},{Meyer},{CatSamba1},{RebRi},{CatSamba2},{Samba}}.

This paper considers an approach that was originally developed in the 
context of atomic physics. It was invented independently by 
Girardeau~\cite{girar1} and Vorob'ev and Khomkin~\cite{russ}. The method has 
been continuously improved throughout the last two decades, and has been used 
by Girardeau and others in several areas of atomic 
physics~\cite{{girar},{coll}}. Although the method shares several properties
with the traditional mappings used in nuclear physics~\cite{KM}, it presents
particularities that make it suitable for hadronic problems as we shall 
discuss shortly ahead. It is based on the ideas of mapping between physical 
and ideal Fock spaces, and has some similarities with the method of Bohm 
and Pines~\cite{bp} to treat collective motions in plasmas. It 
is a generalization of a transformation employed by S. Tani~\cite{tani} in 
1960 to study single-particle scattering by a potential with a bound state. 
Girardeau 
coined the name ``Fock-Tani" representation for this method. 

In the Fock-Tani representation one starts with the Fock representation of the
system using field operators of elementary constituents which satisfy canonical
(anti)commutation relations. Composite-particle field operators  are 
linear combinations of the elementary-particle operators and do not generally
satisfy canonical (anti)commutation relations. ``Ideal" field 
operators acting on an enlarged Fock space are then introduced in close 
correspondence with the composite ones. The enlarged Fock space is a graded 
direct product of the original Fock space and an ``ideal state space". The 
ideal operators correspond to particles with the same quantum numbers of the 
composites; however, they satisfy by definition canonical (anti)commutation 
relations. Next, a given unitary transformation, which transforms the 
single composite states into single ideal states, is introduced. When the 
transformation acts on operators in the subspace of the enlarged Fock space 
which contains no ideal particles, the transformed operators explicitly 
express the interactions of composites and constituents. Application of the 
unitary operator on the microscopic Hamiltonian, or on 
other hermitian operators expressed in terms of the elementary constituent 
field operators such as electroweak currents, gives equivalent operators 
which contain the ideal field operators. The effective Hamiltonians and 
currents in the new representation are hermitian and have a clear physical 
interpretation in terms of the processes they describe. Since all field 
operators in the new representation satisfy canonical (anti)commutation 
relations, the standard methods of quantum field theory can then be readily 
applied.

Several characteristics of the Fock-Tani representation make it suitable to 
the nature of the hadronic problem. It seems to be particularly relevant for 
the building of effective hadronic Hamiltonians in the context of 
effective field theories~\cite{Weff}, as it implements in a certain sense
the ``quasiparticle" method of Weinberg~\cite{quasi}. In Weinberg's 
quasiparticle approach the bound states are redescribed by elementary
particles and, in order not to change the physics of the problem, the 
potential is modified in such a way that it cannot produce these bound states
any longer. In the Fock-Tani representation, as a result of the transformation
of the quark Hamiltonian, the quark-quark interactions become ``weaker", in 
the sense that they describe only quark-quark scattering processes and cannot
produce the hadrons as bound states. The interesting feature of the Fock-Tani
representation is that the change of the potential is the result of the 
unitary transformation that implements the mapping of composite hadrons into
elementary hadrons, while in Weinberg's approach there is some freedom in how 
the potential is modified. Other appealing features of the Fock-Tani 
representation are: (a)~it can be naturally extended to composites
with any number of constituents, not only pairs or triplets of fermions, 
(b)~systems of composite bosons and fermions, i.e. systems containing 
simultaneously mesons and baryons, can be naturally treated in a unified 
manner, and (c)~it can be used with models where creation and annihilation 
of particles play an important role. 

In the following section we review the basic ideas of the Fock-Tani 
representation. We introduce the unitary transformation for meson states of 
a quark-antiquark pair and derive the structure of the quark and antiquark 
operators in the new representation.
Subsection~\ref{sec:FTtransf} contains the necessary extensions to the 
hadronic case of the original developments of Refs.~\cite{{girar},{gilb}}, 
and incorporates the improvements and new aspects developed in 
Refs.~\cite{{gistra},{girlo}}. 
In subsection~\ref{sec:mes-mes} we introduce a model microscopic quark 
Hamiltonian, implement its Fock-Tani transformation, and discuss the physical 
interpretation of the different terms of the transformed Hamiltonian. 
In subsection~\ref{sec:mmpp}, a commonly used quark Hamiltonian is then 
employed to compare the predictions of the Fock-Tani representation with 
other formalisms; in particular we compare results with 
Refs.~\cite{{QBD_mes},{Grefunc}}, and discuss the implications 
of the post-prior symmetry~\cite{schiff} to an application of charmonium 
dissociation in matter~\cite{qua}. In section~\ref{sec:baryons} we 
construct the Fock-Tani transformation for the three-quark baryon systems;
some results hereof were published previously in the form of a
letter~\cite{HKSV}. We perform the Fock-Tani transformation on the quark 
operators in subsection~\ref{sec:operators_bar}. The Hamiltonian that 
describes explicitly the baryon degrees of freedom in the new 
representation is obtained in subsection~\ref{sec:HFT_bar}. 
In subsection~\ref{sec:effnn} we 
present an example of an effective nucleon-nucleon potential as derived from 
a microscopic spin-spin interaction and compare it to the one obtained by 
Barnes and collaborators~\cite{QBD_bar} in the context of the Quark-Born 
diagram (QBD) method. Section~\ref{sec:ortho} contains a technical 
discussion about 
orthogonality corrections for the effective meson-meson and baryon-baryon
Hamiltonian, and presents numerical calculations of their impact on the 
effective interactions. In section~\ref{sec:gen} we discuss the extension of 
the Fock-Tani representation to hadron Fock-space states that are more general
than the quark-antiquark and the three quark states considered in previous 
sections. Such an extension is necessary for treating systems where creation 
and annihilation of particles plays an important role. Our conclusions and 
future perspectives are presented in section~\ref{sec:concl}.

\section{Mapping of mesons}
\label{sec:mesons}

This section reviews the formal aspects of the mapping procedure and
implements it to quark-antiquark meson states. We start with the original 
formulation of Girardeau~\cite{{girar1},{girar}} and include new 
developments and improvements that have occurred since his original work. 
In section~\ref{sec:baryons} we implement the method for three-quark baryonic
states. The more complicated Fock-space states are discussed in 
Section~\ref{sec:gen}.

The starting point of the Fock-Tani method is the definition of single
composite bound states. We write a single-meson state in terms of a meson 
creation operator $M_{\alpha}^{\dagger}$ as
\bea
|\alpha \ran  = M_{\alpha}^{\dagger}|0 \ran ,
\label{1b}
\eea
where the meson creation operator $M_{\alpha}^{\dagger}$ is written
in terms of constituent quark and antiquark creation operators 
$q^{\dagger}$ and $\aq^{\dagger}$,
\bea
M^{\dagger}_{\alpha}= \Phi_{\alpha}^{\mu \nu}
q_{\mu}^{\dagger} {\aq}_{\nu}^{\dagger} .
\label{Mop}
\eea
Here, $\Phi_{\alpha}^{\mu \nu}$ is the Fock-space amplitude of the meson, and 
$|0 \ran$ is the vacuum state, $q_{\mu}|0 \ran=\aq_{\nu}|0\ran=0$.
The index $\alpha$ identifies the quantum numbers of the meson, 
$\alpha=\{$spatial, spin, isospin$\}$. The indices $\mu$ and $\nu$ denote 
the spatial, spin, flavor, and color quantum numbers of the constituent 
quarks. A summation over repeated indices is implied. It is convenient to work
with orthonormalized amplitudes,
\bea
\lan\alpha|\beta\ran= \Phi_{\alpha}^{*\mu \nu} 
\Phi_{\beta}^{\mu \nu}=\delta_{\alpha \beta}.
\label{norm}
\eea
In Section~\ref{sec:mes-mes}, a specific example of a quark model 
is discussed, and the short-hand notation for the labels of the states and 
operators will be spelled out in detail.

The quark and antiquark operators satisfy canonical anticommutation relations,
\beq
\{q_{\mu}, q^{\dagger}_{\nu}\}=
\{\aq_{\mu},\aq^{\dagger}_{\nu}\}=\delta_{\mu \nu},\hspace{1.0cm}
\{q_{\mu}, q_{\nu}\}=\{\aq_{\mu},\aq_{\nu}\}=
\{q_{\mu}, \aq_{\nu}\}= \{q_{\mu}, \aq_{\nu}^{\dagger}\}=0.
\label{qcom}
\eeq
Using these quark anticommutation relations, and the normalization condition
of Eq.~(\ref{norm}), it is easily shown that the meson operators satisfy the 
following noncanonical commutation relations
\beq
[M_{\alpha}, M^{\dagger}_{\beta}]=\delta_{\alpha \beta} - 
\Delta_{\alpha \beta},\hspace{1.5cm}[M_{\alpha}, M_{\beta}]=0,
\label{Mcom}
\eeq
where
\bea
\Delta_{\alpha \beta}= \Phi_{\alpha}^{*{\mu \nu }}
\Phi_{\beta}^{\mu \sigma }\aq^{\dagger}_{\sigma}\aq_{\nu}
+ \Phi_{\alpha}^{*{\mu \nu }}
\Phi_{\beta}^{\rho \nu}q^{\dagger}_{\rho}q_{\mu}.
\label{delta}
\eea
In addition, 
\beq
[q_{\mu},M_{\alpha}^{\dagger}]=\delta_{\mu \mu'}\Phi^{\mu' \nu}_{\alpha}
{\aq}_{\nu}^{\dagger},\hspace{1.0cm}
[{\aq}_{\nu},M_{\alpha}^{\dagger}]=-\delta_{\nu \nu'}\Phi^{\mu \nu'}_{\alpha}
q_{\mu}^{\dagger},\hspace{1.0cm}[q_{\mu},M_{\alpha}]=[{\aq}_{\nu}, M_{\alpha}]=0.
\label{MMq}
\eeq

The presence of the operator $\Delta_{\alpha \beta}$ in Eq.~(\ref{Mcom}) is
due to the composite nature of the mesons. This term enormously complicates 
the mathematical description of processes that involve the
hadron and quark degrees of freedom. The usual field theoretic techniques
used in many-body problems, such as the Green's functions method, Wick's 
theorem, etc, apply to creation 
and annihilation operators that satisfy canonical relations. Similarly, the
non-vanishing of the commutators $[q_{\mu},M_{\alpha}^{\dagger}]$ and 
$[{\aq}_{\nu},M_{\alpha}^{\dagger}]$ is a manifestation of the lack of
kinematic independence of the meson operator from the quark and antiquark
operators. Therefore, the meson operators $M_{\alpha}$ and 
$M_{\alpha}^{\dagger}$ are not convenient dynamical variables to be used. Of 
course, the problem can be formulated in terms of the canonical constituent
field operators only. But then there would be other difficulties, such as 
singularities in certain Green's functions due to the presence of bound 
states.
 
The bound state amplitude $\Phi^{\mu \nu}_{\alpha}$ is obtained from the 
microscopic quark-antiquark Hamiltonian. However, in implementing the
Fock-Tani transformation, the explicit form of the microscopic Hamiltonian is 
not required; the method only requires knowledge of the form of the bound 
states in terms of the constituent operators. Therefore, the following 
discussion is completely general and does not depend on the details of the 
microscopic quark-antiquark interactions of the model. Once the transformation
properties of the quark operators are obtained, the application of the 
transformation on a given microscopic model Hamiltonian gives rise to 
effective Hamiltonians that describe all possible processes involving
quarks and hadrons. 

The idea of the Fock-Tani method is to change representation, such that the
field operators of composite particles are redescribed by field operators 
which satisfy canonical (anti)commutation relations. The complications of the 
composite nature of the hadrons will be shifted to the effective Hamiltonians.
The main features of the Fock-Tani transformation are:

\begin{enumerate}
\item The transformation is performed by a unitary operator $U$, such that 
\beq
|\Omega \ran\longrightarrow |\Omega)=U^{-1}|\Omega \ran,\hspace{1.0cm}
O \longrightarrow O_{\rm FT} = U^{-1}OU.
\label{trans}
\eeq
$|\Omega \ran$ is an arbitrary state vector and $O$ an arbitrary operator, 
both are expressed in terms of the constituent quark and antiquark operators 
$q, q^{\dagger}, \aq, \aq^{\dagger}$ of the original Fock representation. 
$|\Omega)$ and $O_{\rm FT}$ are the corresponding quantities in the new 
representation.  Since $U$ is unitary, scalar products and matrix elements 
are preserved under the change of representation
\beq
\lan  \Omega|\Omega \ran=(\Omega|\Omega),\hspace{1.0cm}
\lan  \Omega|O|\Omega \ran=(\Omega|O_{\rm FT}|\Omega).
\label{preserv}
\eeq
Note that in the new representation, states are represented by rounded, instead
of angular, bras and kets.

\item The transformation is defined such that  a single-meson state 
$|\alpha \ran$ is redescribed by an (``ideal") elementary-meson state by 
\bea
|\alpha \ran\longrightarrow U^{-1}|\alpha \ran\equiv |\alpha)=
m^{\dagger}_{\alpha}|0),
\label{single_mes}
\eea
where $m^{\dagger}_{\alpha}$ an ideal meson creation operator. The ideal
meson operators $m^{\dagger}_{\alpha}$ and $m_{\alpha}$ satisfy,
by definition, canonical commutation relations
\beq
[m_{\alpha}, m^{\dagger}_{\beta}]=\delta_{\alpha \beta} ,
\hspace{1.5cm}[m_{\alpha}, m_{\beta}]=0.
\label{mcom}
\eeq
The state $|0)$ is the vacuum of both $q$ and $m$ degrees of freedom in the 
new representation. A precise definition of $|0)$ is given later in the text.
In addition, in the new representation the quark and antiquark operators 
$q^{\dagger}$, $q$, $\aq^{\dagger}$ and $\aq$ are kinematically independent of 
the $m^{\dagger}_{\alpha}$ and $m_{\alpha}$
\bea
[q_{\mu},m_{\alpha}]=[q_{\mu},m^{\dagger}_{\alpha}]=[\aq_{\mu},m_{\alpha}]=
[\aq_{\mu},m^{\dagger}_{\alpha}]=0 .
\label{indep_mes}
\eea

\item A multi-meson state $|\alpha_1, \cdots, \alpha_n \ran$, constructed from 
mutually non-overlapping and well-separated wave packets, is transformed into 
a multi-ideal-meson state $|\alpha_1, \cdots, \alpha_n)$
\bea
|\alpha_1, \cdots, \alpha_n \ran \rightarrow U^{-1}|\alpha_1, \cdots, 
\alpha_n \ran
&=&|\alpha_1, \cdots, \alpha_n)
=m^{\dagger}_{\alpha_1} \cdots m^{\dagger}_{\alpha_n}|0).
\label{nonovlp}
\eea
This is particularly important for meson-meson scattering processes, where
asymptotic states are non-overlapping. However, Eq.~(\ref{nonovlp}) is not 
true in the general case of multi-meson states describing a dense system of 
mesons, where one expects considerable overlap among the mesons.

\item Given a microscopic quark-antiquark Hamiltonian operator, the Fock-Tani
transformed Hamiltonian can generally be written as
\bea
H\rightarrow H_{\rm FT} = U^{-1}HU \equiv H_{\rm FT}^{(0)}+V_{\rm FT}.
\label{FTham}
\eea
$H_{\rm FT}^{(0)}$ is the ``non-interacting" part; it contains the 
single-quark part of the original Hamiltonian, and a single-meson part which 
describes the free propagation of the composite mesons. $V_{\rm FT}$ is the 
interacting part, responsible for all possible interactions among the 
composites and the quarks. It describes only the ``true" interaction 
processes, the binding effects of the quarks and antiquarks into mesons are
contained in $H_{\rm FT}^{(0)}$. 
\end{enumerate}

The unitary operator $U$ that implements the transformation,
Eq.~(\ref{single_mes}), is constructed as follows: let the physical Fock 
space be denoted by~$\cal F$. This is the space of states generated by all
linear combinations of states constructed from the quark and antiquark 
operators $q^{\dagger}$ and ${\aq}^{\dagger}$,
\bea
q_{\mu_1}^{\dagger} \cdots q_{\mu_l}^{\dagger} 
{\aq}_{\nu_1}^{\dagger} \cdots {\aq}_{\nu_m}^{\dagger}|0 \ran,
\label{linq}
\eea
with $l$ and $m$ arbitrary. The quark and antiquark operators satisfy
canonical anticommutation relations, Eq.~(\ref{qcom}), and $|0 \ran$ is the
vacuum state defined as above.

Now, an independent Hilbert space $\cal H$, the ``ideal hadron
space" is defined, as the space of all linear combinations of states 
constructed from the ideal meson operators $m^{\dagger}$,
\bea
m_{\alpha_1}^{\dagger} \cdots m_{\alpha_n}^{\dagger}|0 )_{\cal H},
\label{linm}
\eea
where $|0 )_{\cal H}$ is the vacuum of ${\cal H}$, that is, 
$m_{\alpha}|0 )_{\cal H}=0$.

Next, the ``ideal state space" $~{\cal I}$ is defined as the graded direct 
product of the Fock space  $\cal F$ and the ideal hadron space $\cal H$:  
$~{\cal I}={\cal F} \times {\cal H}$. By definition, the quark and ideal 
meson operators are 
kinematically independent and therefore satisfy Eq.~(\ref{indep_mes}) 
on $\cal I$. The commutation relations given in Eqs.~(\ref{qcom}) and 
(\ref{Mcom}), initially   defined on $\cal F$, as well as those in 
Eq.~(\ref{mcom}), initially defined on $\cal H$, are also valid for $\cal I$. 

The vacuum state of space $\cal I$ is the direct product of the vacua of 
$\cal F$ and $\cal H$. To simplify the notation, the symbol $|0)$ is used to 
denote the product $|0)\equiv |0 \ran \times |0 )_{\cal H}$.
Therefore, $|0)$ is the vacuum of both the quark and ideal hadron degrees of
freedom, $q_{\mu}|0)=\aq_{\mu}|0)=m_{\alpha}|0)=0$. The quark operators act 
on $|0 \ran$, and the ideal operators on $|0 )_{\cal H}$.

The objective is to establish an one-to-one correspondence between the 
composite states in $\cal F$ and the ideal states of a subspace of $\cal I$. 
To that extent, we note that in $\cal I$ there is a subspace which is 
isomorphic to the original Fock space $\cal F$, namely, the space 
${\cal I}_{0}$ consisting of those states $|\Omega \ran$ with no ideal mesons,
\bea
m_{\alpha}|\Omega \ran=0.
\label{constr}
\eea
This indicates that in ${\cal I}_{0}$ the ideal mesons are 
``redundant modes", in the language of the Bohm and Pines method~\cite{bp}. 
As we shall discuss shortly, Eq.~(\ref{constr}) is 
a constraint condition to ensure that there will be no double counting of 
degrees of freedom. This condition plays a role similar to the ``negative
particles'' in the CPR of Ref.~\cite{WuFeng}.  In the 
original formulation of the method~\cite{girar1}~\cite{girar}, the change of 
representation is performed by the operator
\bea
U=\exp\left(-{\pi \over 2}F \right),
\label{uandf}
\eea
where the generator of the transformation $F$ is given by 
\bea
F = m^{\dagger}_{\alpha}M_{\alpha}-
M^{\dagger}_{\alpha}m_{\alpha},
\label{f_mes}
\eea
with $M^{\dagger}_{\alpha}$ given by Eq.~(\ref{Mop}). As before, a summation 
over repeated indices is implied.  We note that $U$ acts on 
$\cal I$ and cannot be defined on $\cal F$. However, it is defined on 
${\cal I}_{0}$, which is isomorphic to $\cal F$. The image 
${\cal F_{\rm FT}}=U^{-1}{\cal I}_{0}$ of ${\cal I}_{0}$
is the subspace of $\cal I$ which consists of all states $|\Omega)$ in the
new representation, related to the states $|\Omega \ran$ of ${\cal I}_{0}$ by
\bea
|\Omega)=U^{-1}|\Omega \ran.
\label{relsta}
\eea
In the subspace ${\cal F}_{\rm FT}$, Eq.~(\ref{constr}) is transformed to
\bea
U^{-1}m_{\alpha}U|\Omega )=0.
\label{constransf}
\eea 
Therefore, any calculation in the original Fock space is equivalent to a 
calculation in the Fock-Tani space $\cal F_{\rm FT}$ when 
Eq.~(\ref{constransf}) is satisfied . It is not difficult 
to prove~\cite{gilb} that the transformation implemented by such an operator 
$U$ does indeed have the characteristics 1) to 4) as discussed above.

The great advantage of working in $\cal F_{\rm FT}$ is that all
creation and annihilation operators satisfy canonical commutation
relations. The transformed operators 
$O_{\rm FT}=U^{-1}OU$ give rise, in general, to an infinite series.
This series physically represents an expansion in the ``degree of overlap'' 
of the composites in the system. For the derivation of an effective two-meson 
interaction, only a few terms in the series are necessary. A potential
complication is the constraint equation, Eq.~(\ref{constransf}), called the 
``subsidiary condition". For scattering problems, where one starts with the 
proper definition of the asymptotic states, Eq.~(\ref{constransf}) is 
trivially satisfied. It is also satisfied in the particularly important case 
of the many ideal mesons state $m_{\alpha_1}^{\dagger} \cdots 
m_{\alpha_n}^{\dagger}|0)$~\cite{girlo}. 

In the next subsection we present the derivation of the transformed quark 
operators.


\subsection{Transformation of the quark operators}
\label{sec:FTtransf}

The next step is to obtain the transformed operators in the new representation.
The basic operators of the model, such as the Hamiltonian, electromagnetic
currents, etc,  are expressed in terms of the quark operators. Therefore, in
order to obtain the basic operators of the model in the new representation, 
the transformed quark operators are needed
\bea
q_{\rm FT}=U^{-1}\, q \,U,\hspace{1.0cm}{\aq}_{\rm FT}=U^{-1}\, {\aq}\, U .
\eea
The evaluation of such expressions by direct multi-commutators is tremendously
difficult, it involves the summation of infinite series and cannot in general
be expressed in a closed form. However, Girardeau~\cite{girar} suggests the 
use of an ``equations of motion" technique, which consists of the following. 
For any operator $\calo$, a ``time-dependent" operator is defined as
\bea
\calo (t) = \exp{(-tF)}\, \calo\, \exp{(tF)},
\label{ot}
\eea
where $t$ is a real ``time" parameter. Differentiating the above equation
with respect to $t$, an equation of motion for $\calo (t)$ is obtained,
\bea
\frac{d\calo (t)}{dt}=[\calo (t), F],
\label{eqmot_mes}
\eea
with the ``initial condition"
\bea
\calo (0)=0.
\label{bc}
\eea
Therefore, the transformed operators are obtained from the solution of 
Eqs.~(\ref{eqmot_mes})~and~(\ref{bc}) by setting $t=-\pi/2$ at the
end
\bea
\calo_{\rm FT}=U^{-1}\, \calo \, U=\calo (-\pi/2).
\eea

The equations of motion for the quark operators $q$ and $\aq$ can be obtained
by making use of Eq.~(\ref{MMq}), which leads to
\bea
&&{dq_{\mu}(t) \over dt } = \left[q_{\mu}(t), F\right]
=-\delta_{\mu \mu_1} \Phi^{\mu_1 \nu}_{\alpha}
\aq^{\dagger}_{\nu}(t)m_{\alpha}(t),
\label{eqq}\\
&&{d\aq_{\nu}(t) \over dt} = \left[{\aq}_{\nu}(t), F\right]
=\delta_{\nu \nu_1} \Phi^{\mu \nu_1}_{\alpha} q^{\dagger}_{\mu}(t)
m_{\alpha}(t).
\label{eqqb}
\eea
Since these equations involve $m_{\alpha}(t)$, the equation of motion for 
$m_{\alpha}(t)$ is also needed
\bea
{dm_{\alpha}(t) \over dt} = [m_{\alpha}(t),F] = M_{\alpha}(t).
\label{eqm}
\eea
The system of equations is closed with 
\bea
{dM_{\alpha}(t) \over dt} = [O_{\alpha}(t),F]  
= -\left[\delta_{\alpha \beta}-\Delta_{\alpha \beta}(t)\right]m_{\beta}(t).
\label{eqM}
\eea
Eqs.~(\ref{eqq})-(\ref{eqM}) (together with their hermitian conjugates)
form a set of nonlinear ordinary differential equations. Obviously this set of
equations is as complicated to solve as the evaluation of the multicommutators
discussed above. However, these equations can be solved in a 
straightforward way by iteration. Starting from a ``zero-order" approximation,
where the overlap among the mesons is neglected, terms of the same ``power" 
in the bound state amplitudes $\Phi_{\alpha}$ and 
$\Phi^{*}_{\alpha}$ are collected. For each operator the expansions are
then written,
\beq
q_{\mu}(t)=\sum_{i=1}^{\infty}q_{\mu}^{(i)}(t),\hspace{0.5cm}
\aq_{\mu}(t)=\sum_{i=1}^{\infty}\aq_{\mu}^{(i)}(t),\hspace{0.5cm}
m_{\alpha}(t)=\sum_{i=1}^{\infty}m_{\alpha}^{(i)}(t),\hspace{0.5cm}
M_{\alpha}(t)=\sum_{i=1}^{\infty}M_{\alpha}^{(i)}(t) ,
\label{expans}
\eeq
where the superscript $i$ denotes the $i$-th power of $\Phi_{\alpha}$ and 
$\Phi^{*}_{\alpha}$ in the series. In order to have a consistent power 
counting, the implicit $\Phi_{\alpha}$ and  $\Phi^{*}_{\alpha}$ entering via 
Eq.~(\ref{Mop}) are not counted~\cite{girar}. In order to derive an effective 
meson-meson interaction, the expansion up to the third order in the wave 
functions~\cite{girar} is needed. Therefore, we will explicitly 
derive the transformed operators up to the third order. 

In the zeroth-order approximation, the effects of the meson structure are 
neglected. This amounts to neglect the terms
$\Delta_{\alpha \beta}$ and $\Phi_{\alpha}^{\mu \nu}$ of 
Eqs.~(\ref{eqq})~,~(\ref{eqqb})~and~(\ref{eqM}):
\beq
{d q_{\mu}^{(0)}(t)\over dt} = 0,
\hspace{0.5cm}{d \aq_{\nu}^{(0)}(t)\over dt}=0,\hspace{0.5cm} 
{d M_{\alpha}^{(0)}(t)\over dt}=-m_{\alpha}^{(0)}(t),
\hspace{0.5cm}
{d m_{\alpha}^{(0)}(t)\over dt}=M_{\alpha}^{(0)}(t).
\label{0eqs}
\eeq
Using the initial condition, Eq.~(\ref{bc}), the zero-order solutions are 
found to be:
\bea
&&q^{(0)}_{\mu}(t)=q_{\mu},
\hspace{0.5cm} 
\aq^{(0)}_{\nu}(t)=\aq_{\nu}, \nn\\
&&m^{(0)}_{\alpha}(t)=m_{\alpha}\cos t + M_{\alpha}\sin t, 
\hspace{0.5cm}
M^{(0)}_{\alpha}(t)=M_{\alpha}\cos t - m_{\alpha}\sin t.
\label{sol0}
\eea
At this zeroth-order approximation, the transformed mesons behave 
like elementary bosons, and the transformation of $M_{\alpha}$ to $m_{\alpha}$ 
has the interpretation of a rotation of $-\pi/2$ in the space ${\cal I}$. 

The first-order equations are given by
\bea
&&{d q^{(1)}_{\mu}(t) \over dt}=-\delta_{\mu \mu_1}
\Phi^{\mu_1 \nu_{1}}_{\alpha}\aq^{\dagger (0)}_{\nu_{1}}(t)m^{(0)}_{\alpha}(t),
\hspace{0.5cm}{d \aq^{(1)}_{\nu}(t) \over dt}=\delta_{\nu \nu_1} 
\Phi^{\mu_{1}\nu_1}_{\alpha}q^{\dagger (0)}_{\mu_{1}}(t)m^{(0)}_{\alpha}(t), 
\nn\\
&&{d M_{\alpha}^{(1)}(t)\over dt}=-m_{\alpha}^{(1)}(t), 
\hspace{1.0cm}
{d m_{\alpha}^{(1)}(t)\over dt}=M_{\alpha}^{(1)}(t).
\eea
Since the initial conditions were assigned to the zero-order terms, one
obtains
\bea
q^{(i)}_{\mu}(t=0)=\aq^{(i)}_{\mu}(t=0)=m^{(i)}_{\alpha}(t=0)=
M^{(i)}_{\alpha}(t=0)=0,
\hspace{0.1cm} {\text{for}}\hspace{0.1cm} i\geq 1 .
\label{leq1}
\eea
Therefore, the solutions to the first-order equations can be written as
\bea
&&q^{(1)}_{\mu}(t) = -\,\delta_{\mu \mu_1}\Phi^{\mu_1 \nu_{1}}_{\alpha}
\aq^{\dagger }_{\nu_{1}} \left[m_{\alpha} \sin t + M_{\alpha} 
\left(1 - \cos t \right)\right],\\
&&\aq^{(1)}_{\nu}(t) = +\,\delta_{\nu \nu_1} \Phi^{\mu_{1}\nu_1 }_{\alpha}
q^{\dagger }_{\mu_{1}} \left[ m_{\alpha} \sin t + M_{\alpha} 
\left(1 - \cos t\right)
\right],\\
&&m^{(1)}_{\alpha}(t)=0,\hspace{1.0cm}M^{(1)}_{\alpha}(t)=0.
\label{sol1}
\eea

The iterative procedure can be continued to higher orders in a straightforward
way. However, the solutions of second and higher orders will give rise to 
secular terms; i.e. terms which involve polynomials in $t$, in addition to 
trigonometric functions in $t$. Among other problems, the secular terms 
introduce the familiar post-prior discrepancies in the analysis of scattering 
and reactive processes. The origin of the secular terms is the asymmetry of 
the equations of motion for $m_{\alpha}(t)$ and $M_{\alpha}(t)$, 
Eqs.~(\ref{eqm})~and~(\ref{eqM}); the term proportional to 
$\Delta_{\alpha \beta}$ spoils the symmetry. The solution to the problem was 
recently found by Girardeau and Straton~\cite{gistra}. The solution consists 
in adding to $F$ a term depending on $\Delta_{\alpha \beta}$, such that the 
equations become symmetrical. 

Here, a more systematic and elegant, although equivalent procedure
of Lo and Girardeau~\cite{girlo} is followed. The generator of the 
transformation $F$ is generalized to
\bea
F=m^{\dagger}_{\alpha}O_{\alpha}-O^{\dagger}_{\alpha}
m_{\alpha},
\label{newF}
\eea
where the operator $O_{\alpha}$ is a function of the quark and antiquark 
operators, and are chosen such that 
\bea
[O_{\alpha},O^{\dagger}_{\beta}]= \delta_{\alpha\beta},\hspace{1.5cm}
[O_{\alpha},O_{\beta}]=[O^{\dagger}_{\alpha},O^{\dagger}_{\beta}]=0.
\label{comO}
\eea
The consequence of this is that the equations for $m_{\alpha}$ and $O_{\alpha}$
are manifestly symmetric,
\beq
{d m_{\alpha}(t)\over dt}=[m_{\alpha}(t),F]=O_{\alpha}(t),\hspace{1.0cm}
{d O_{\alpha}(t)\over dt}=[O_{\alpha}(t),F]=-m_{\alpha}(t),
\label{solm&O}
\eeq
and their solutions involve only trigonometric functions of $t$, 
\beq
m_{\alpha}(t)=O_{\alpha} \sin t + m_{\alpha} \cos t , \hspace{1.0cm}
O_{\alpha}(t)=O_{\alpha} \cos t - m_{\alpha} \sin t.
\label{solOm}
\eeq

The operators $O_{\alpha}$ are obtained order-by-order in a expansion in 
power of the bound state wave functions, as follows. In the zeroth-order, it 
is clear that 
\bea
O_{\alpha}^{(0)}=M_{\alpha}.
\label{O0_mes}
\eea 
This certainly satisfies Eq.~(\ref{comO}) to zeroth order, and  reproduces 
the original results at the zeroth and first orders. Since there is no first 
order contribution to $O_{\alpha}$, because $\Delta_{\alpha\beta}$,
which is the term to be canceled in the commutation relation, is already of 
the second order. Therefore,
\bea
O_{\alpha}=M_{\alpha}+O_{\alpha}^{(2)},
\eea
where $O_{\alpha}^{(2)}$ must be chosen such that
\bea
[O_{\alpha},O^{\dagger}_{\beta}]= \delta_{\alpha\beta} + 
{\cal O}(\Phi^{3}).
\label{ft4}
\eea
The appropriate choice is easily checked to be:
\bea
O_{\alpha}^{(2)}=\frac{1}{2}\Delta _{\alpha \beta}M_{\beta}.
\label{O2_mes}
\eea
Following the same course, the appropriate third order 
operator $O_{\alpha}^{(3)}$ is discovered to be given by
\bea
O_{\alpha}^{(3)}=-\frac{1}{2}M^{\dagger}_{\beta}[\Delta_{\beta\gamma},
M_{\alpha}]M_{\gamma}.
\label{O3_mes}
\eea
Note that a sum over repeated indices is implied in 
Eqs.~(\ref{O2_mes})~and~(\ref{O3_mes}).

Using the new generator $F$, Eq.~(\ref{newF}), the second order equations for 
the quark and antiquark operators are obtained in a straightforward way. 
They are given by
\bea
d q^{(2)}_{\mu}(t)\over dt &=& -\, \delta_{\mu \mu_1}\left[
\Phi^{\mu_1\nu_1}_{\alpha}\aq^{\dagger(1)}_{\nu_1}(t)m^{(0)}_{\alpha}(t)
+ \frac{1}{2}\,\Phi^{\ast\mu_2\nu_1}_{\alpha}  
\Phi^{\mu_1\nu_1}_{\beta} m^{\dagger(0)}_{\alpha}(t)q^{(0)}_{\mu_2}(t)
M^{(0)}_{\beta}(t)\right.\nn\\
&-& \left.\frac{1}{2}\, \Phi^{\ast\mu_2\nu_1}_\alpha\Phi^{\mu_1\nu_1}_\beta 
M^{\dagger(0)}_{\alpha}(t)q^{(0)}_{\mu_2}(t)m^{(0)}_{\beta}(t)\right],\nn \\
{d \aq^{(2)}_{\nu}(t)\over dt}&=& \delta_{\nu \nu_1}\left[ 
\Phi^{\mu_1\nu_1}_{\alpha} 
q^{\dagger(0)}_{\mu_1}(t)m^{(0)}_{\alpha}(t) 
+ \frac{1}{2}\Phi^{\ast\mu_{1}\nu_2}_{\alpha} \Phi^{\mu_1\nu_1}_{\beta}
m^{\dagger(0)}_{\alpha}(t)M^{(0)}_{\beta}(t)
\aq^{(0)}_{\nu_2}(t) \right. \nn\\
&-&\left. \frac{1}{2}\Phi^{\ast\mu_1\nu_2}_{\alpha}
\Phi^{\mu_1\nu_1}_{\beta}M^{\dagger(0)}_{\alpha}(t)
m^{(0)}_{\beta}(t)\aq^{(0)}_{\nu_2}(t)\right].
\label{eqb2}
\eea
The integration of these equations leads to the following expressions
\bea
q^{(2)}_{\mu}(t)&=&\delta_{\mu \mu_1}\frac{1}{2}\Phi^{\ast\mu_2\nu_1}_{\alpha}
\Phi^{\mu_1\nu_1}_{\beta}\left[
m^{\dagger}_{\alpha}M_{\beta} \sin t \cos t - m^{\dagger}_{\alpha}m_{\beta}\sin^2t
- M^{\dagger}_{\alpha}M_{\beta}\left(1-\cos t\right)^2 \right.\nn \\ 
&-&\left. M^{\dagger}_{\alpha}m_{\beta} \left(2-\cos t\right) \sin t
\right]q_{\mu_2}\\
\aq^{(2)}_{\nu}(t)&=&\delta_{\nu\nu_1}\frac{1}{2}\Phi^{\ast\mu_1\nu_2}_{\alpha}
\Phi^{\mu_1\nu_1}_{\beta}\left[m^{\dagger}_{\alpha}M_{\beta}
\sin t \cos t-m^{\dagger}_{\alpha}m_{\beta}\sin^{2}t-M^{\dagger}_{\alpha}
M_{\beta}\left(1-\cos t\right)^2\right.\nn\\
&-&\left. M^{\dagger}_{\alpha}m_{\beta} \left(2-\cos t\right) \sin t
\right]\aq_{\nu_2}.
\eea

In the same way, one can derive the third order equations
\bea
{dq^{(3)}_{\mu}(t) \over dt}&=&-\,\delta_{\mu\mu_1}\frac{1}{2}
\Phi^{\mu_1\nu_{1}}_{\alpha}
\left\{
2 \left[\aq^{\dagger (2)}_{\nu_{1}}(t)
m^{(0)}_{\alpha}(t) +\aq^{\dagger (0)}_{\nu_{1}}(t)m^{(2)}_{\alpha}(t) \right]
-\aq^{\dagger(0)}_{\nu_1}(t)
\Delta_{\alpha\beta}(t)m^{(0)}_{\beta}(t) 
\right.
\nn \\
&+&\Phi^{\ast\mu_2\nu_1}_{\beta}
\left[m^{\dagger(0)}_{\beta}(t)q^{(1)}_{\mu_2}(t)M^{(0)}_{\alpha}(t)
-M^{\dagger(0)}_{\beta}(t)q^{(1)}_{\mu_2}(t) 
m^{(0)}_{\alpha}(t)\right] 
\nn \\ 
&+&\left.
\Phi^{\ast\rho\nu_1}_{\beta}\Phi^{\rho\sigma}_{\gamma}
M^{\dagger(0)}_{\beta}(t) \aq^{\dagger(0)}_{\sigma}(t)
\left[ M^{(0)}_{\gamma}(t)m^{(0)}_{\alpha}(t)+
M^{(0)}_{\alpha}(t)m^{(0)}_{\gamma}(t)\right]\right\},\label{eqq3}\\
{d\aq^{(3)}_{\nu}(t) \over dt}&=& \delta_{\nu \nu_1}\frac{1}{2}
\Phi^{\mu_{1}\nu_1}_{\alpha} 
\left\{ 2\left[q^{\dagger (0)}_{\mu_{1}}(t) m^{(2)}_{\alpha}(t)  
+q^{\dagger (2)}_{\mu_{1}}(t)m^{(0)}_{\alpha}(t)\right] 
+q^{\dagger(0)}_{\mu_1}(t)\Delta_{\alpha\beta}(t)m^{(0)}_{\beta}(t)
\right. \nn \\
&+&\Phi^{\ast\mu_1\sigma}_{\beta}
\left[m^{\dagger(0)}_{\beta}(t)\aq^{(1)}_{\sigma}(t)M^{(0)}_{\alpha}(t)-
M^{\dagger(0)}_{\beta}(t)\aq^{(1)}_{\sigma}(t)
m^{(0)}_{\alpha}(t)\right]
\nn \\
&-&\left.
\Phi^{\ast\mu_1\sigma}_{\beta}
\Phi^{\rho\sigma}_{\gamma}
M^{\dagger (0)}_{\beta}(t)q^{\dagger(0)}_{\rho}(t)
\left[M^{(0)}_{\alpha}(t)m^{(0)}_{\gamma}(t)+
M^{(0)}_{\gamma}(t)m^{(0)}_{\alpha}(t)\right]\right\}.
\label{eqqb3}
\eea
In order to integrate these equations $m_{\alpha}^{(2)}(t)$ is needed,
which can be obtained from Eqs.~(\ref{solOm})~and~(\ref{O2_mes})
\bea
m_{\alpha}^{(2)}(t)= O_{\alpha}^{(2)} \sin t + m_{\alpha} \cos t 
=\frac{1}{2}\Delta _{\alpha \beta}M_{\beta} \sin t + m_{\alpha} \cos t.
\label{solm2}
\eea
Using this in Eqs.~(\ref{eqq3})~and~(\ref{eqqb3}) above, and integrating, 
the third order quark and antiquark operators are obtained
\bea
q^{(3)}_{\mu}(t)&=& \delta_{\mu \mu_1}\frac{1}{2}\Phi^{\mu_1\nu_1}_{\alpha}
 \left\{
\Phi^{\ast\rho\nu_1}_{\beta}\Phi^{\rho\sigma}_{\gamma}
\aq^{\dagger}_{\sigma}
\left[
m^{\dagger}_{\beta}m_{\alpha}m_{\gamma} \sin^{3}t + 
M^{\dagger}_{\beta}M_{\alpha}m_{\gamma}\sin t \cos^{2} t 
\right.\right. \nn\\ 
&+& M^{\dagger}_{\beta}m_{\alpha}M_{\gamma} 
\left(2-\cos t - \sin^{2} t\right) \sin t 
-\left(M^{\dagger}_{\beta}m_{\alpha}m_{\gamma} + 
m^{\dagger}_{\beta}M_{\alpha}m_{\gamma}\right)
\sin^2 t \cos t \nn\\
&+&m^{\dagger}_{\beta}m_{\alpha}M_{\gamma}
\left(1-\cos t\right) \sin^2 t 
+M^{\dagger}_{\beta}M_{\alpha}M_{\gamma}
\left(1+\cos^2 t\right)\left(1-\cos t\right)\nn\\
&+&\left.\left. m^{\dagger}_{\beta}M_{\alpha}M_{\gamma} 
\left(\cos t - 1\right)\cos t \sin t\right]+
\aq^{\dagger}_{\nu_{1}}\Delta_{\alpha\beta}
\left[2M_{\beta}\left(\cos t-1\right)-m_{\beta}\sin t\right]\right\}
\label{solq3}\nn\\
\aq^{(3)}_{\nu}(t)&=&-\,\delta_{\nu \nu_1}\frac{1}{2}
\Phi^{\mu_1\nu_1}_{\alpha} 
\left\{
\Phi^{\ast\mu_1\sigma}_{\beta}\Phi^{\rho_{1}\sigma}_{\gamma}
q^{\dagger}_{\rho_1}\left[
m^{\dagger}_{\beta}m_{\alpha}m_{\gamma}\sin^{3}t +
M^{\dagger}_{\beta}M_{\alpha}m_{\gamma} \sin t \cos^{2} t
\right. \right. \nn\\ 
&+& M^{\dagger}_{\beta}m_{\alpha}M_{\gamma} 
\left(2-\cos t - \sin^{2} t\right) \sin t
-\left(M^{\dagger}_{\beta}m_{\alpha}m_{\gamma}+m^{\dagger}_{\alpha}M_{\beta}
m_{\gamma}\right) \sin^{2} t \cos t \nn\\
&+& m^{\dagger}_{\beta}m_{\alpha}M_{\gamma}
\left(1-\cos t\right)\sin^2 t
+M^{\dagger}_{\beta}M_{\alpha}M_{\gamma}
\left(1 + \cos^2 t \right)\left(1-\cos t\right)\nn\\
&+& \left.\left. m^{\dagger}_{\beta}M_{\alpha}M_{\gamma} 
\left(\cos t - 1\right)\cos t \sin t \right]+
q^{\dagger}_{\mu_{1}}\Delta_{\alpha\beta}
\left[2M_{\beta}\left(1-\cos t\right)+m_{\beta}\sin t\right]
\right\} .
\label{solqb3}
\eea

This completes the derivation of the transformed operators up to the third 
order. The formulae above provide the starting point for the construction of 
effective Hamiltonians for a variety of processes involving mesons and quarks,
such as an  effective meson-meson Hamiltonian. There are higher-order terms 
that provide orthogonality corrections to the lowest-order ones. These are 
rather trivial to obtain and so the discussion of this material is defered to 
section~\ref{sec:ortho}.


\subsection{Effective Meson Hamiltonian}
\label{sec:mes-mes}

The Fock-Tani Hamiltonian $H_{\rm FT}$ is obtained from the microscopic
quark-antiquark Hamiltonian by the application of the unitary operator $U$,
as indicated in Eq.~(\ref{FTham}). Therefore, the structure of the microscopic
Hamiltonian must be specified. We consider a microscopic Hamiltonian of the 
general form
\bea
H &=& T\left(\mu\right)q^{\dagger}_{\mu} q_{\mu} + T\left(\nu\right)
\aq_{\nu}^{\dagger}\aq_{\nu} \nn\\
&+& V_{q\aq}(\mu\nu;\sigma\rho)q^{\dagger}_{\mu}\aq^{\dagger}_{\nu}
\aq_{\rho}q_{\sigma} + \frac{1}{2} V_{qq}(\mu\nu;\sigma\rho)
q^{\dagger}_{\mu}q^{\dagger}_{\nu}q_{\rho}q_{\sigma}
+\frac{1}{2}V_{\aq\aq}(\mu\nu;\sigma\rho)\aq^{\dagger}_{\mu}
\aq^{\dagger}_{\nu}\aq_{\rho}\aq_{\sigma}.
\label{qHamilt}
\eea
The convention of a summation over repeated indices is again assumed.
We note that a great variety of quark-model Hamiltonians used in the 
literature can be written in such a form. However, at this point of the 
discussion we have not included in Eq.~(\ref{qHamilt}) terms such as 
pair-creation, which are of the form $\aq^{\dagger}q^{\dagger}q^{\dagger}q$. 
It will become clear from the discussion below that such terms can be handled 
with no additional problem. We also defer to section~\ref{sec:gen} the 
discussion of the more complicated structures of the Fock-space decomposition 
of meson states. 

The transformation is implemented by transforming each quark and antiquark 
operator in Eq.~(\ref{qHamilt}). In free space, the single-meson Fock-space 
amplitudes $\Phi^{\mu\nu}_{\alpha}$ of Eq.~(\ref{Mop}) satisfy the following 
equation
\beq
H(\mu\nu; \mu'\nu')\Phi_{\alpha}^{\mu' \nu'}=\epsilon_{[\alpha]} 
\Phi_{[\alpha]}^{\mu \nu},
\label{Schro}
\eeq
where $H(\mu\nu; \mu'\nu')$ is the Hamiltonian matrix
\bea
H(\mu\nu; \mu'\nu')=\delta_{\mu[ \mu']} \delta_{\nu[\nu']} 
\left[T([\mu']) + T([\nu'])\right]+ 
V_{q\aq}(\mu\nu; \mu' \nu') ,
\label{Schroperat}
\eea
and $\epsilon_{[\alpha]}$ is the total energy of the meson (center-of-mass
energy plus internal energy). We use the convention that there is no sum 
over repeated indices inside square brackets. We note that in order to 
implement the unitary transformation, there is no fundamental reason to
use Fock-space amplitudes $\Phi^{\mu\nu}_{\alpha}$ that satisfy the free-space 
equation, Eq.~(\ref{Schro}) above. However, as we shall discuss shortly,
such a choice facilitates many formal manipulations of the equations. 

The transformed Hamiltonian $H_{\rm FT}=U^{-1}HU$ contains the same information
as the original microscopic Hamiltonian, and as such describes all possible 
processes involving mesons and quarks. Such processes include two-body 
quark-quark, meson-quark, and meson-meson interactions, as well as processes
that involve more than two particles (quarks and mesons). As we already 
discussed previously, the unitary transformation can be evaluated, in general, 
only to a finite order in the meson Fock-space amplitude. This implies that 
only a limited class of processes are contained in the transformed Hamiltonian,
evaluated up to a certain order. Up to the order that the quark operators are 
evaluated in this paper, it is possible to obtain an effective 
Hamiltonian that describes few-particle interactions. In the following,
the discussion is specialized to the two-body processes, which are the 
quark-quark, meson-quark, and meson-meson processes. The derivation of the 
transformed Hamiltonian follows the same path as in 
Ref.~\cite{girar}~\cite{girlo} for the case of atomic physics, where the 
bound states are of the same structure as our meson states. We have therefore 
skiped some intermediate steps, and refer the reader to noted references for 
details. In section~\ref{sec:baryons} baryons are considered, and since the 
method has not been implemented before for three-particle bound states, the 
transformation of the Hamiltonian is presented with greater detail than in the 
present section. 

The structure of the transformed Hamiltonian, $H_{\rm FT}=U^{-1}HU$ is of 
the following form
\bea
H_{\rm FT}=H_{q} + H_{m} + H_{m q} ,
\label{separation}
\eea
where the first term involves only quark operators, the second one involves 
only ideal meson operators, and $H_{m q}$ involves quark and meson operators.
A similar separation of the Hamiltonian is also explicitly obtained in the 
approach of Ref.~\cite{WuFeng} using the CPR. In Eq.~(\ref{separation}), the 
quark Hamiltonian $H_{q}$ has an identical structure to the microscopic quark 
Hamiltonian, Eq.~(\ref{qHamilt}), except that the term corresponding to the 
quark-antiquark interaction is modified such that the new interaction does not
produce quark-antiquark bound states. The new quark-antiquark interaction 
becomes modified as
\bea
V_{q\bar{q}} &=&\left[V_{q\aq}(\mu \nu; \sigma \rho) -
\Delta(\mu \nu; \mu' \nu')H(\mu' \nu'; \sigma \rho) -
H(\mu \nu; \sigma' \rho')\Delta(\sigma' \rho'; \sigma \rho)\right.\nn \\ 
&+& \left. \Delta(\mu \nu; \mu' \nu')H(\mu' \nu'; \sigma'\rho')
\Delta(\sigma' \rho'; \sigma \rho)\right]\, 
q^{\dagger}_{\mu}\aq^{\dagger}_{\nu} \aq_{\rho} q_{\sigma},
\label{modqaq}
\eea
where $\Delta(\mu \nu; \mu' \nu')$ is the ``bound state kernel"
\bea
\Delta(\mu \nu; \mu' \nu')=\sum_{\alpha}\Phi^{\mu \nu}_{\alpha}
\Phi^{*\mu' \nu'}_{\alpha}.
\label{kernel}
\eea
For the sake of clarity, here, and in a few subsequent formulas, we explicitly
write the sum over repeated quantum numbers of mesons. One property of the 
bound state kernel we repeatedly make use of is 
\beq
\Delta(\mu \nu; \mu' \nu')\Phi^{\mu'\nu'}_{\alpha}=\Phi^{\mu\nu}_{\alpha},
\label{propDelta}
\eeq
which follows from the orthonormalization of the $\Phi$'s, Eq.~(\ref{norm}).
In the case that the $\Phi^{\mu\nu}_{\alpha}$'s are stationary states of the 
microscopic Hamiltonian, i.e. they are solutions of
Eq.~(\ref{Schro}), the quark-antiquark interaction term is modified as
\bea
V_{q\aq}(\mu \nu ; \sigma \rho) \longrightarrow 
V_{q\aq}(\mu \nu ; \sigma \rho) - \sum_{\alpha}
\epsilon_{\alpha}\Phi^{\ast \mu \nu}_{\alpha} \Phi^{\sigma \rho}_{\alpha},
\label{modV}
\eea
where $\epsilon_{\alpha}$ is the meson total energy [see Eq.~(\ref{Schro})].
It is not difficult to show that the spectrum of the modified quark 
Hamiltonian, $H_{q}$, is positive semi-definite and hence has no bound 
states~\cite{girar}. This feature is exactly the same as in Weinberg's 
quasiparticle method~\cite{quasi}. In Weinberg's approach, the bound states 
are redescribed by elementary or ideal particles, and in order not to change 
the physics of the problem, the potential is modified in such a way that it 
cannot produce these bound states anymore. In the present formalism, this 
feature happens automatically.

The term involving only ideal meson operators has the general form
\bea
H_{m}= T_{m} + V_{mm},
\label{hmm}
\eea
where $T_{m}$ is the single-meson term,
\bea
T_{m}=\sum_{\alpha\beta} \Phi^{*\mu\nu}_{\alpha} H(\mu\nu;\mu'\nu')
\Phi^{\mu'\nu'}_{\beta}m^{\dagger}_{\alpha} m_{\beta},
\label{Tm}
\eea
and $V_{mm}$ is the effective meson-meson interaction
\bea
V_{mm}=\frac{1}{2} \sum_{\alpha \beta \gamma \delta}
V_{mm}(\alpha \beta; \gamma \delta)m^{\dagger}_{\alpha}m^{\dagger}_{\beta}
m_{\delta}m_{\gamma}.
\label{mesonV}
\eea
For later convenience, we have divided the potential into direct (dir), 
exchange (exc), and intra-exchange (int) parts as 
\begin{equation}
V_{mm}(\alpha\beta;\gamma\delta)=V_{mm}^{dir}(\alpha\beta;\gamma\delta)
+V_{mm}^{exc}(\alpha\beta;\gamma \delta)
+V_{mm}^{int}(\alpha\beta;\gamma\delta),
\label{Vmm}
\end{equation}
where each of these is given by
\bea
V_{mm}^{dir}(\alpha \beta;\gamma \delta)&=& 
2\Phi_{\alpha}^{* \mu \sigma}\Phi_{\beta}^{* \rho \nu}
V_{q\aq}(\mu \nu ;\mu'\nu')\Phi_{\delta}^{\rho \nu'}
\Phi_{\gamma}^{\mu' \sigma}+\Phi_{\alpha}^{* \mu \sigma}
\Phi_{\beta}^{* \rho \nu}V_{qq}(\mu \rho ;\mu' \rho')
\Phi_{\delta}^{\rho' \nu}\Phi_{\gamma}^{\mu' \sigma}\nn\\
&+&\Phi_{\alpha}^{* \mu \sigma}\Phi_{\beta}^{* \rho \nu}
V_{\aq\aq}(\sigma \nu ;\sigma' \nu')
\Phi_{\delta}^{\rho \nu'}\Phi_{\gamma}^{\mu \sigma'},
\label{Vdir_mes}
\eea
\bea
V_{mm}^{exc}(\alpha \beta;\gamma \delta)&=& -\frac{1}{2}\left[
\Phi_{\alpha}^{* \mu \nu}\Phi_{\beta}^{* \rho \sigma}
V_{q\aq}(\mu \nu ;\mu' \nu')\Phi_{\delta}^{\mu' \sigma}
\Phi_{\gamma}^{\rho \nu'} +
\Phi_{\alpha}^{* \rho \sigma}\Phi_{\beta}^{* \mu \nu}
V_{q\aq}(\mu \nu ;\mu' \nu')\Phi_{\delta}^{\rho \nu'}
\Phi_{\gamma}^{\mu' \sigma}\right.\nn\\
&+&\Phi_{\alpha}^{* \mu \sigma}\Phi_{\beta}^{* \rho \nu}
V_{q\aq}(\mu \nu ;\mu' \nu')\Phi_{\delta}^{\mu' \nu'}
\Phi_{\gamma}^{\rho \sigma} +
\Phi_{\alpha}^{* \rho \nu}\Phi_{\beta}^{* \mu \sigma}
V_{q\aq}(\mu \nu ;\mu' \nu')\Phi_{\delta}^{\rho \sigma}
\Phi_{\gamma}^{\mu' \nu'}\nn\\
&+&\left. 2 \Phi_{\alpha}^{* \mu \sigma}\Phi_{\beta}^{* \rho \nu}
V_{qq}(\mu \rho ;\mu'\rho')\Phi_{\delta}^{\mu' \nu}
\Phi_{\gamma}^{\rho' \sigma}+
2\Phi_{\alpha}^{* \mu \sigma}\Phi_{\beta}^{* \rho \nu}
V_{\aq\aq}(\sigma \nu ;\sigma' \nu')\Phi_{\delta}^{\mu \nu'}
\Phi_{\gamma}^{\rho \sigma'}\right],
\label{Vexch_mes}
\eea
\bea
V_{mm}^{int}(\alpha \beta;\gamma \delta)&=&  -\frac{1}{2} \left[
\Phi_{\alpha}^{* \mu \nu}\Phi_{\beta}^{* \rho \sigma}
H(\mu \nu;\mu'\nu')\Phi_{\delta}^{\mu' \sigma}\Phi_{\gamma}^{\rho \nu'} + 
\Phi_{\alpha}^{* \rho \sigma}\Phi_{\beta}^{* \mu \nu}H(\mu \nu ;\mu' \nu')
\Phi_{\delta}^{\rho \nu'}\Phi_{\gamma}^{\mu' \sigma}\right.\nn\\
&+&\left. \Phi_{\alpha}^{* \mu \sigma}\Phi_{\beta}^{* \rho \nu}
H(\mu \nu ;\mu' \nu')\Phi_{\delta}^{\mu' \nu'}
\Phi_{\gamma}^{\rho \sigma} + \Phi_{\alpha}^{* \rho \nu}
\Phi_{\beta}^{* \mu \sigma}H(\mu \nu ;\mu' \nu')
\Phi_{\delta}^{\rho \sigma}\Phi_{\gamma}^{\mu' \nu'}\right].
\label{Vintra}
\eea
In general, the effective meson-meson potential is non-local, even when
the microscopic interaction is a local one. The nonlocality is, of course, due 
to the extended structure of the mesons, and the size of the nonlocality is 
typically of the order of the size of the meson. 

In section~\ref{sec:ortho} we will present a particularly interesting property
of $V_{mm}^{int}$, namely, that it is canceled at the lowest order by 
orthogonality corrections. These orthogonality corrections have the same 
origin as the subtraction of the bound states from the microscopic 
quark-antiquark interaction, and in the present case they appear as powers of 
the bound state kernel $\Delta$, defined in Eq. (\ref{kernel}). As a result of
this cancelation, the effective meson-meson potential is then given by the 
first two terms in Eq.~(\ref{Vmm}).

The term $ H_{m q}$ describes a variety of processes involving mesons and
quarks, such as quark-meson scattering, meson breakup into quarks, etc. 
One of these terms represents the two-meson breakup into two quarks and two 
antiquarks. The structure of such a term is
\bea
H_{mm \rightarrow q\aq q\aq} =\frac{1}{2}V_{\alpha\beta}(\mu\nu ; \sigma\rho)\,
q_{\mu}^{\dagger} \aq_{\nu}^{\dagger} q_{\rho}^{\dagger} \aq_{\sigma}^{\dagger}
m_{\beta} m_{\alpha} , 
\label{break}
\eea
where
\bea
V_{\alpha \beta}(\mu \nu ;\sigma \rho)=V_{qq}(\mu \rho ; \mu' \sigma')
\Phi_{\alpha}^{\sigma' \sigma} \Phi_{\beta}^{\mu' \nu}
+V_{\aq \aq}(\nu \sigma ; \nu' \sigma')\Phi_{\alpha}^{\mu \sigma'}
\Phi_{\beta}^{\rho \nu'}+2V_{q \aq}(\mu \sigma ; \mu' \nu')
\Phi_{\alpha}^{\mu' \nu} \Phi_{\beta}^{\rho \nu'}.
\label{Vbreak}
\eea
When this term is iterated, together with its hermitian conjugate 
$H_{q\aq q\aq\rightarrow mm }= H_{mm \rightarrow q\aq q\aq}^{\dagger}$, which
represents the recombination of two quarks and two antiquarks into two mesons,
the following picture of a physical process occurs: (1) in the collision
process the two mesons breakup into two quarks and two antiquarks under the 
action of $H_{mm \rightarrow q\aq q\aq}$, (2) then the quarks and antiquarks 
propagate freely, and (3) due to the confining forces they recombine into 
mesons under the action of $H_{q\aq q\aq\rightarrow mm }$. 
This mechanism is competitive with the one of the same second order Born 
approximation, where in the intermediate state the quarks and antiquarks 
remain bound into mesons. It would be interesting to investigate possible 
observable differences between these two mechanisms in a truly confining 
model, since this would allow to discriminate genuine quark effects from 
mesonic ones in the scattering process. 

Before proceeding to the next subsection, we notice that the Fock-space hadron
amplitudes used to define the unitary operator $U$, in principle, do not need 
to be precisely the exact bound states of the original microscopic quark 
Hamiltonian. One might start with given ``bare" amplitudes, with parameters to 
be determined at a later stage. The physical asymptotic states can always be 
obtained as ``dressed" states that follow from the ``bare" ones, using the
effective Hamiltonians. That is, it is not necessary to obtain first 
the bound states of the microscopic Hamiltonian, and then perform the mapping. 
There might exist situations where the use of a set of ``bare" amplitudes to 
generate the mapping is more convenient than using the complete set of exact 
bound states of the microscopic Hamiltonian.

In the next subsection, the application of the formalism for 
meson-meson scattering is illustrated through simple examples. We have
specialized our applications in order to compare our results with existing 
calculations from the literature.

\subsection{Meson-meson scattering and post-prior symmetry}
\label{sec:mmpp}

In this subsection, the results are compared with other methods presented in 
the literature. We have compared our meson-meson results with the ones obtained
recently in Refs.~\cite{QBD_mes}~\cite{Grefunc}~\cite{qua}. In order to do so,
we use, as do the above references, a Fermi-Breit Hamiltonian, 
which includes the kinetic term, a spin-spin part of the one-gluon-exchange, 
and a gaussian potential for the confinement of the quarks and antiquarks.
This choice is usually made for reasons of simplicity, as it allows one to 
perform almost all of the calculations analytically. We do not discuss 
the limitations and problems of such a model, since our aim is to compare 
results from the literature, and as such, we need to use the same model. In 
order to make the presentation self-contained, we begin with a collection of  
formulas relating the potential to scattering amplitudes and cross sections. 
We also present a brief discussion of the post-prior discrepancy, which is a 
problem that plagues composite-particle formalisms. We then show explicitly, 
through a numerical example, the effect of the breaking of this symmetry and 
discuss how the present formalism solves the problem.

The exact two-meson scattering and reaction amplitudes are given by the 
$T$-matrix elements in the Fock-Tani representation as
\beq
T(\alpha\beta;\gamma\delta)(z)=<\alpha\beta|T(z)|\gamma\delta>
=(\alpha\beta|T_{FT}(z)|\gamma\delta),
\label{Telem}
\eeq
where the asymptotic final and initial meson states
\bea
|\alpha\beta)&=&U^{-1}M^{\dagger}_{\alpha}M^{\dagger}_{\beta}|0\!>=
m^{\dagger}_{\alpha}m^{\dagger}_{\beta}|0\!),\\
|\gamma\delta)&=&U^{-1}M^{\dagger}_{\gamma}M^{\dagger}_{\delta}|0\!>=
m^{\dagger}_{\gamma}m^{\dagger}_{\delta}|0\!)\,,
\eea
are eigenstates of the operator $T_m$ of Eq. (\ref{Tm}). Note that for 
asymptotic scattering states, the unitary transformation is trivial, as 
discussed at the beginning of this section. The transition operator 
$T_{FT}(z)$ satisfies a Lippmann-Schwinger equation
\beq
T_{FT}(z)=V_{FT}+V_{FT}G_0(z)T_{FT}(z)\,,
\label{LSeq}
\eeq
where $V_{FT}$ is the interacting part of the Fock-Tani Hamiltonian, and
$G_0(z)$ is the noninteracting propagator
\beq
G_0(z)=(z-H^{(0)}_{FT})^{-1}\,.
\label{G0}
\eeq
Here, $H^{(0)}_{FT}$ corresponds to the single-particle part of $H_{FT}$ 
\beq
H^{(0)}_{FT}=\sum_{\mu}T(\mu)q^{\dagger}_{\mu}q_{\mu} 
+\sum_{\nu}T(\nu)\aq^{\dagger}_{\nu}\aq_{\nu} +
\sum_{\alpha\beta} \Phi^{*\mu\nu}_{\alpha} H(\mu\nu;\mu'\nu')
\Phi^{\mu'\nu'}_{\beta}m^{\dagger}_{\alpha} m_{\beta}.
\label{H0FT}
\eeq
To first order, only the meson-meson part of the effective potential $V_{FT}$ 
contributes to the scattering matrix, since the asymptotic states contain 
only ideal mesons, and thus
\beq
T(\alpha'\beta';\gamma'\delta')(z)=(\! \alpha'\beta'|V_{mm}|
\gamma'\delta' \!)\,. 
\eeq

The differential cross-section for the process $i \rightarrow f$ can be given 
in terms of the relativistic invariant matrix element, ${\cal M}_{fi}$, as
\beq
\frac{d\sigma_{fi}(s,t,u)}{dt} = \frac{1}{64\pi s}
\frac{1}{P^2(s)}{|{\cal M}_{fi}(s,t,u)|}^{2}\,,
\label{scd}
\eeq
where $P(s)$ is the relative three-momentum of the initial mesons in 
the center-of-mass frame and $s,t,u$ are the Mandelstam variables.
Due to translational invariance, the T-matrix element can be written as a
momentum conservation delta-function, times a Born-order matrix element, 
$h_{fi}$
\beq
T(\alpha\beta;\gamma\delta)(z)=\delta^{(3)}({\bP}_{f}-{\bP}_{i})
h_{fi}, 
\label{hfi}
\eeq
where ${\bP}_{f}={\bp}_{\alpha}+{\bp}_{\beta}$ and
${\bP}_{i}={\bp}_{\gamma}+{\bp}_{\delta}$ are the final and initial 
momenta of the two-meson system. The invariant matrix element is then given by
\beq
{\cal M}_{fi}=\frac{1}{{(2\pi)}^{3}}\prod_{n=1}^{4}{[2{(2\pi)}^{3}
E_{n}]}^{\frac{1}{2}}h_{fi},
\eeq
and the total cross-section is obtained by integrating Eq.~(\ref{scd}) over t,
\beq
\sigma_{fi}(s)=\int_{t_{-}}^{t_{+}}dt\,\frac{d\sigma_{fi}(s,t,u)}
{dt},
\label{Sigma}
\eeq
where $t_{-}$ and  $t_{+}$ are the minimum and maximum transfer momenta . 

Consider the rearrangement collision between composite particles $\gamma$ and
$\delta$, resulting in particles $\alpha$ and $\beta$, 
$\gamma + \delta \rightarrow \alpha +\beta $.
In this type of reaction, the constituents of the colliding systems can be
redistributed during the process, and thus, the reaction may result in the same
original states (elastic scattering) or in different states (inelastic
scattering). The full Hamiltonian which describes the system is usually split
into free and interaction parts as $H=H^{0}+V$. Because of the internal 
degrees of freedom, this decomposition is not unique~\cite{schiff}. One could,
for instance, choose a decomposition in which $H^{0}$ is diagonal on the 
initial or final asymptotic states, 
\bea
H=H^{0}_{i}+V_{i}=H^{0}_{f}+V_{f}\,.
\eea
Generally, $H^{0}_{i} \neq H^{0}_{f}$ and $V_{i} \neq V_{f}$, since the
initial and final rearrangement channels can be different. Thus, the T-matrix
element that describes the scattering can be written in either the prior form
or in the post form, with the interaction respectively in either the initial 
channel or in the final channel 
\beq
T(\alpha\beta;\gamma\delta)_{prior}=
<\! \Psi^{-}_{\alpha\beta}|V_{i}|\phi_{\gamma\delta} \!>,\hspace{1.0cm}
T(\alpha\beta;\gamma\delta)_{post}=<\! \phi_{\alpha\beta}|V_{f}
|\Psi^{+}_{\gamma\delta} \!>.
\label{Tprior-Tpost}
\eeq
The initial and final free asymptotic states, $\phi_{\gamma\delta}$ and
$\phi_{\alpha\beta}$, are eigenstates of the free Hamiltonians, $H^{0}_{i}$
and $H^{0}_{f}$. The outgoing and incoming exact scattering states,
$\Psi^{+}_{\gamma\delta}$ and $\Psi^{-}_{\alpha\beta}$, are eigenstates of
the full Hamiltonian and satisfy Lippmann-Schwinger equations, 
\beq
\Psi^{+}_{\gamma\delta}= \phi_{\gamma\delta}+G_{0}^{+}V_{i} 
\Psi^{+}_{\gamma\delta},\hspace{1.0cm}
\Psi^{-}_{\alpha\beta}= \phi_{\alpha\beta}+G_{0}^{-}V_{f} 
\Psi^{-}_{\alpha\beta}\,,
\eeq
where $G_{0}^{+}$ and $G_{0}^{-}$ are the retarded and advanced free-particle
Green's functions given by 
\beq
G_{0}^{+}=\lim_{\epsilon \rightarrow 0^{+}}[E-H^{0}_{i}+i\epsilon]^{-1},
\hspace{1.0cm}
G_{0}^{-}=\lim_{\epsilon \rightarrow 0^{+}}[E-H^{0}_{f}-i\epsilon]^{-1}.
\eeq
It can be proved that the prior and post forms are strictly equivalent on the
energy shell~\cite{joachin}. Thus, the exact T-matrix is symmetrical under
the interchange of initial and final states: 
\bea
T(\alpha\beta;\gamma\delta)=
<\! \Psi^{-}_{\alpha\beta}|V_{i}|\phi_{\gamma\delta} \!>=<\! 
\phi_{\alpha\beta}|V_{f}|\Psi^{+}_{\gamma\delta} \!>.
\eea

The first-Born approximation consists in replacing
$\Psi^{-}_{\alpha\beta}$ by $\phi_{\alpha\beta}$ and
$\Psi^{+}_{\gamma\delta}$ by $\phi_{\gamma\delta}$ in 
Eq.~(\ref{Tprior-Tpost}). Thus, the prior and post forms of the first-Born 
T-matrix elements are given by
\beq
T_{B}(\alpha\beta;\gamma\delta)_{prior}=<\! \phi_{\alpha\beta}|V_{i}|
\phi_{\gamma\delta} \!>,\hspace{1.0cm}
T_{B}(\alpha\beta;\gamma\delta)_{post}=<\! \phi_{\alpha\beta}|V_{f}|
\phi_{\gamma\delta} \!>.
\eeq
These two expressions are equal only if the free asymptotic states, 
$\phi_{\gamma\delta}$ and $\phi_{\alpha\beta}$, are exact eigenstates of 
their respective free Hamiltonians. When approximate wave functions are 
used for the bound states, one generally finds different values for the prior 
and post matrix elements. This difference is known as the ``post-prior'' 
discrepancy. It appears in many derivations of scattering amplitudes 
for reactions involving composite particles, since the exact bound state wave 
functions are usually not exactly known. 

The lack of this symmetry will be shown not to be of importance for the 
``symmetric'' initial and final states, as in the case of 
$\pi + \pi \rightarrow \pi + \pi$~\cite{QBD_mes}~\cite{Grefunc}; however, it 
is of importance for asymmetric cases like 
$J/\Psi + \pi \rightarrow D$-mesons~\cite{qua}. Of course, one way to 
cure the problem, as done by Swanson in the first reference of 
Ref.~\cite{QBD_mes}, is to use the exact eigenstates of the free Hamiltonian. 
However, it might occur in a realistic situation that the problem is 
sufficiently complicated, and that an exact solution is unobtainable. Within 
the Fock-Tani formalism, a post-prior symmetrical effective Hamiltonian is
automatically attained. As a result of this, a first-order T-matrix element 
that is free of post-prior discrepancy is obtained, even for approximate 
eigenstates of the ``free" Hamiltonian.

Let us start specifying the quark-quark, antiquark-antiquark, and 
quark-antiquark terms in Eq.~(\ref{qHamilt}) that are used in
Refs.~\cite{QBD_mes}~\cite{Grefunc}~\cite{qua}. Spelled out in full detail, 
these are given by
\bea
V_{qq}&=&\frac{1}{2}\int \frac{d^3{\bk}_1 d^3{\bk}_2 d^3{\bk}_3d^3{\bk}_4}
{(2\pi)^3}\,\delta_{f_1 f_3}\delta_{f_2 f_4} \,
\left(\frac{\lambda^a}{2}\right)^{c_1 c_3}
\left(\frac{\lambda^a}{2}\right)^{c_2 c_4}\nn\\ 
&&\hspace{0.25cm}\times\, U_{s_1s_3;s_2s_4}(\bk_1-\bk_3)\,
q^{c_1\dagger}_{s_1 f_1}(\bk_1)\,q^{c_2\dagger}_{s_2 f_2}(\bk_2)\,
q^{c_4}_{s_4 f_4}(\bk_4)\,q^{c_3}_{s_3 f_3}(\bk_3),
\label{Vqq}\\
V_{\aq\aq}&=&\frac{1}{2}\int \frac{d^3{\bk}_1 d^3{\bk}_2 d^3{\bk}_3d^3{\bk}_4}
{(2\pi)^3} \, \delta_{f_1 f_3}\delta_{f_2 f_4} \,
\left(\frac{\lambda^{aT}}{2}\right)^{c_1 c_3}
\left(\frac{\lambda^{aT}}{2}\right)^{c_2 c_4}\nn\\ 
&&\hspace{0.25cm}\times\, U_{s_1s_3;s_2s_4}(\bk_1-\bk_3)\,
\aq^{c_1\dagger}_{s_1 f_1}(\bk_1)\,\aq^{c_2\dagger}_{s_2 f_2}(\bk_2)\,
\aq^{c_4}_{s_4 f_4}(\bk_4)\,\aq^{c_3}_{s_3 f_3}(\bk_3),\label{Vaqaq}\\
V_{q\aq}&=&-\int \frac{d^3{\bk}_1 d^3{\bk}_2 d^3{\bk}_3d^3{\bk}_4}
{(2\pi)^3}\,\delta_{f_1 f_3}\delta_{f_2 f_4}\,
\left(\frac{\lambda^a}{2}\right)^{c_1 c_3}
\left(\frac{\lambda^{aT}}{2}\right)^{c_2 c_4}\,\\
&&\hspace{0.25cm}\times\,U_{s_1s_3;s_2s_4}(\bk_1-\bk_3)\,
q^{c_1\dagger}_{s_1 f_1}(\bk_1)\,\aq^{c_2\dagger}_{s_2 f_2}(\bk_2)\,
\aq^{c_4}_{s_4 f_4}(\bk_4)\,q^{c_3}_{s_3 f_3}(\bk_3),
\label{Vqaq}
\eea
where $\lambda^a/2$ for quark and $-\lambda^{a T}/2$ 
($T$ means transpose) for antiquarks, are the $SU(3)$ color matrices. The 
potential functions are given by
\beq
U_{s_1s_3;s_2s_4}({\bq})=\delta^{(3)}({\bk}_1 + {\bk}_2 - {\bk}_3 - {\bk}_4)\,
\left[U_{s_1s_3;s_2s_4}^{conf}({\bq})
+ U_{s_1s_3;s_2s_4}^{ss}({\bq})\right]\,, 
\label{mic}
\eeq
with
\bea
&&U_{s_1s_3;s_2s_4}^{conf}({\bq})=\delta_{s_1s_3} \delta_{s_2s_4} \left[
V_0 \left(\frac{4\pi}{\chi}\right)^{\frac{3}{2}} e^{-{\bq}^2/\chi}+
C\right] \,,
\\
&&U_{s_1s_3;s_2s_4}^{ss}({\bq})=-\frac{8\pi\alpha_s}{3m_1 m_2}
\left(\frac{\sigma^i}{2}\right)_{s_1s_3} 
\left(\frac{\sigma^i}{2}\right)_{s_2s_4}\,,
\label{mic3}
\eea
where $m_1$ and $m_2$ are the constituent quark and antiquark masses, 
$\sigma^i$ ($i=1,2,3$) are the Pauli spin matrices, $\alpha_{s}$ is the 
QCD structure constant, $V_0$ and $\chi$ are the parameters of the gaussian
potential, and $C$ is a constant to adjust the lower part of the spectrum.

The explicit form of the creation operator for a composite meson is 
\bea
M^{\dagger}_{CSF}({\bp}) = \sum_{csf} {\cal C}^{c_1c_2}_C 
\chi^{s_1s_2}_S {\cal F}^{f_1f_2}_F 
\int d{\bk}_1 d{\bk}_2 \,\Phi_{\bp}({\bk}_1, {\bk}_2)
q^{\dagger}_{c_1s_1f_1}({\bk}_1) \bar q^{\dagger}_{c_2s_2f_2}({\bk}_2),
\eea
where ${\cal C}_C$, $\chi_{S}$, and ${\cal F}_F$ are respectively the color,
spin and flavor Clebsch-Gordan coefficients, and $\Phi_{\bp}$ is the spatial
amplitude, an S-wave gaussian given by
\beq
\Phi_{\bp}^{\bk_1\bk_2}=
\delta^{(3)}(\bp-\bk_1-\bk_2)\left(\frac{b^2}{\pi}\right)^{\frac{3}{4}}
e^{-b^2 \bar{\bk}^2/2 },
\label{quark_mes_wf}
\eeq
where $\bar{\bk}=\eta\bk_1-(1-\eta)\bk_2$, 
with $\eta=m_2/(m_1+m_2)$. The gaussian parameter $b$ 
is related to the r.m.s. radius of the meson by $<r^2>=\sqrt{3/2}\,b$.

Let us now compare our results with the ones from Ref.~\cite{QBD_mes}, which
uses the QBD method, and those from Ref.~\cite{Grefunc}, which 
uses a Green's functions method. These references consider $I=2$ meson-meson 
scattering, and consider $m_1=m_2=m_q$. In the present formalism the 
effective meson-meson potential for this channel is given by
\bea
V_{mm}&=&\frac{1}{2}\int d^3{\bP}\, d^3{\bP}'\,d^3{\bp}\,d^3{\bp}'\,
\delta^{(3)}({\bP}'-{\bP})\,\left[V_{conf}({\bp},{\bp}')
+ V_{ss}({\bp},{\bp}')\right]\nn\\
&&\hspace{0.13cm}\times\, m^{\dagger}_{\lambda_1}({\bP}'/2+{\bp}')\,
m^{\dagger}_{\lambda_2}({\bP}'/2-{\bp}')\,m_{\lambda_4}({\bP}/2-{\bp})\,
m_{\lambda_3}({\bP}/2+{\bp}),
\label{VmmFT}
\eea
where $\lambda_i$ ($i=1,\cdots 4$) are the meson isospin quantum numbers 
(in this example they correspond to the quantum numbers of $\pi^+$ ), and 
$V_{conf}$ and $V_{ss}$ are respectively the contributions from the confining 
and the spin-spin interactions, given by
\begin{eqnarray}
V_{conf}(\bp,\bp')&=&\frac{2V_0b^3}{9}
\left(\frac{z}{2\pi}\right)^{\frac{3}{2}}
\Biggl\{\frac{8}{\left[z(4+3x)\right]^{\frac{3}{2}}} \Biggl[
e^{-b^2 \left({\bp'}^2+y{\bp}^2\right)/4} +
e^{-b^2\left(y{\bp'}^2+{\bp}^{2}\right)/4}\Biggr] \nn\\
&&\hspace{0.13cm}-\, e^{-b^2\left({\bp'}^2+
{\bp}^2\right)/4}\Biggl[e^{xz b^2\left({\bp'}+{\bp}\right)^2/8}+
e^{xz b^2\left({\bp'}-{\bp}\right)^2/8}\Biggr]\Biggr\} ,
\label{conf}\\
\nn\\
V_{ss}({\bp},{\bp}')&=&\frac{\kappa_{ss}}{6}
\Biggl\{\frac{8}{3\sqrt{3}}
\Biggl[ e^{-b^2\left({\bp'}^2+{\bp}^2/3\right)/4}+
e^{-b^2\left({\bp}^2+{\bp'}^2/3\right)/4}\Biggr]\nn\\
&&\hspace{0.13cm}+\,e^{-b^2\left({\bp'}-{\bp}\right)^{2}/8}+
e^{-b^2\left({\bp'}+{\bp}\right)^{2}/8}\Biggr\},
\label{hyp}
\end{eqnarray}
where
\bea
\kappa_{ss}=\frac{\alpha_s}{3m_q^2\,\pi^2}
\hspace{.5cm},\hspace{.5cm}
x=\frac{b^2\chi}{2}
\hspace{.5cm},\hspace{.5cm}
y=\frac{(4+x)}{(4+3x)}
\hspace{.5cm},\hspace{.5cm}
z=\frac{1}{(1+x)}.
\label{kss}
\eea 

Comparing our expressions with Eqs.~(25)~and~(26) of Ref.~\cite{Grefunc},
two differences are immediately noticed: (1) we have extra terms because in 
the meson-meson potential we have included the contributions coming from 
quark-quark and antiquark-antiquark interactions, whereas in 
Ref.~\cite{Grefunc} only the quark-antiquark was used, and (2) the different 
overall numerical coefficient is due to the fact that we have removed
the color factors $\lambda^a \lambda^a/4$ from the interactions.  It is also
noticeable from the above expressions that the effective meson-meson potential
is symmetrical under the exchange of initial and final states 
(${\bp} \leftrightarrow {\bp}'$). 

Considering only the spin-spin term in Eq.~(\ref{mic}), and writing the 
scattering amplitude $h_{fi}$ defined in Eq.~(\ref{hfi}) in terms of
Mandelstam variables, the following Born-order on-shell matrix 
element is obtained:
\bea
h_{fi}=\frac{\kappa_{ss}}{3}
\left[\frac{16}{3\sqrt{3}}e^{-b^2(s-4M_{mes}^{2})/12}
+e^{b^2t/8}+e^{b^2u/8}\right],
\eea
where
\beq
M_{mes} = 2m_q + \frac{3}{2b^2m_q}
-\frac{8\alpha_{s}}{3m_{q}^{2}b^2\sqrt\pi}.
\label{Energ}
\eeq
This result is identical to the one obtained in Ref.~\cite{QBD_mes} with the 
QBD method.  

Now we compare results for an ``asymmetrical" reaction, namely, charmonium 
dissociation by inelastic scattering on pions~\cite{qua}
\beq
J/{\psi}(^{3}1S_{1}) + \pi(^{1}1S_{0})\rightarrow D(1S)+{\bar D}(1S).
\eeq
The $J/{\psi}$ mesons are composites of a heavy quark and a heavy antiquark 
pair, denoted by $(Q{\bar Q})$, and the pions are composites of a light quark 
and a light antiquark, denoted by $(q\aq)$. The final mesons $D,{\bar D}$ 
are composites of a $(q{\bar Q})$ or a $(Q{\aq})$ pair and can be either in 
the fundamental $D,{\bar D}(^{1}1S_{0})$ or in the excited 
$D^{*},{\bar D}^{*}(^31S_{1})$ states. By momentum conservation, the only
 allowed final meson states are the excited ones. The threshold for each 
process is given by the mass difference
\beq
\Delta M=(M_{D}+M_{\bar D})-(M_{J/ {\psi}}+M_{\pi})\,.
\eeq
The cross-section for this process was calculated by Martins et al.~\cite{qua}
using the QBD. They employ the microscopic interaction given 
by Eqs.~(\ref{mic})-(\ref{mic3}), but the confinement 
gaussian potential is not considered in the $qq$ and $\aq\aq$ channels, and the
spin-spin term is neglected in all channels. In Ref.~\cite{qua} the 
prior potential matrix elements are considered. In order to compare results, 
only the confinement contribution to the invariant matrix 
element is considered, and $V_0$ is multiplied by a numerical constant in 
order to compensate for the presence of Gell-Mann matrices in our 
Eq.~(\ref{Vqaq}). 

In our formalism, the invariant matrix element can be written as 
\beq
{\cal M}_{fi}^{conf} = \frac{1}{2}\left[({\cal M}_{q\aq}^{conf})_{prior} +
({\cal M}_{q\aq}^{conf})_{post}\right],
\label{prior+post}
\eeq
where the prior and post invariant matrix elements are obtained respectively 
from the prior and post effective potentials (note that because of the
Gell-Mann matrices, the $V^{dir}_{mm}$ part in Eq.~(\ref{Vmm}) does not
contribute)
\bea
&&V_{mm}(\alpha \beta;\gamma \delta)_{prior}= -\Bigl[
\Phi_{\alpha}^{* \mu \nu}\Phi_{\beta}^{* \rho \sigma}
V_{q\aq}(\mu \nu ;\mu' \nu')\Phi_{\delta}^{\mu' \sigma}
\Phi_{\gamma}^{\rho \nu'} +
\Phi_{\alpha}^{* \rho \sigma}\Phi_{\beta}^{* \mu \nu}
V_{q\aq}(\mu \nu ;\mu' \nu')\Phi_{\delta}^{\rho \nu'}
\Phi_{\gamma}^{\mu' \sigma}\nn\\
&&\hspace{1.0cm}+\,\Phi_{\alpha}^{* \mu \sigma}\Phi_{\beta}^{* \rho \nu}
V_{qq}(\mu \rho ;\mu'\rho')\Phi_{\delta}^{\mu' \nu}
\Phi_{\gamma}^{\rho' \sigma} +
\Phi_{\alpha}^{* \mu \sigma}\Phi_{\beta}^{* \rho \nu}
V_{\aq\aq}(\sigma \nu ;\sigma' \nu')\Phi_{\delta}^{\mu \nu'}
\Phi_{\gamma}^{\rho \sigma'}\Bigr],\label{prior}
\eea
\bea
&&V_{mm}(\alpha \beta;\gamma \delta)_{post} = - \Bigl[
\Phi_{\alpha}^{* \mu \sigma}
\Phi_{\beta}^{* \rho \nu}V_{q\aq}(\mu \nu ;\mu' \nu')
\Phi_{\delta}^{\mu' \nu'}\Phi_{\gamma}^{\rho \sigma} 
+ \Phi_{\alpha}^{* \rho \nu}
\Phi_{\beta}^{* \mu \sigma}
V_{q\aq}(\mu \nu ;\mu' \nu')\Phi_{\delta}^{\rho \sigma}
\Phi_{\gamma}^{\mu' \nu'}\nn\\
&&\hspace{1.0cm}+\,\Phi_{\alpha}^{* \mu \sigma}
\Phi_{\beta}^{* \rho \nu}
V_{qq}(\mu \rho ;\mu'\rho')\Phi_{\delta}^{\mu' \nu}
\Phi_{\gamma}^{\rho' \sigma}
+\Phi_{\alpha}^{* \mu \sigma}\Phi_{\beta}^{* \rho \nu}
V_{\aq\aq}(\sigma \nu ;\sigma' \nu')\Phi_{\delta}^{\mu \nu'}
\Phi_{\gamma}^{\rho \sigma'}\Bigr]
\label{post}.
\eea
The term $({\cal M}_{q\aq}^{conf})_{prior}$ is equal to the 
matrix element $a{\bar b}+{\bar a}b$ described in Appendix C of 
Ref.~\cite{qua}. Using the notation of Ref.~\cite{qua}, one can write
\bea
({\cal M}_{q\aq}^{conf})_{prior}={\cal M}_{fi}^{QBD}=
{\cal M}_{a{\bar b}}^{np}+{\cal M}_{{\bar a}b}^{np} . 
\eea
That is, the prior form of our matrix element is one-half of 
${\cal M}_{fi}^{QBD}$ used in Ref.~\cite{qua}. The post form of the invariant 
amplitude is given by
\bea
({\cal M}_{q\aq}^{conf})_{post}={\cal N}\frac{4}{9}
\frac{V_0}{{(2\pi)}^3}{(\lambda_{QQ}\lambda_{qq})}
^{\frac{3}{4}}{(32\pi)}^{\frac{3}{2}}x^{\frac{3}{2}}e^{-\alpha|{\bP}|^{2}}
\left(\frac{1}{\Lambda_1}
e^{-\beta_1|{\bP'}|^2}+\frac{1}{\Lambda_{2}}
e^{-\beta_2|{\bP'}|^2}\right)\,,
\eea
where $x=1/(2\chi)$, $\alpha=\lambda_{Qq}$ and
\bea
&&|{\bP}|^2=\frac{1}{4s}\left[{(s-(M_{J/{\psi}}^2+M_{\pi}^2)}^{2}-
4M_{J/{\psi}}^2M_{\pi}^2\right],\nn\\
&&|{\bP'}|^2=\frac{1}{4s}\left[{(s-(M_{D}^2+M_{\bar D}^2)}^{2}-
4M_{D}^2 M_{\bar D}^2\right],\nn\\
&&{\cal N}=\frac{{(2\pi)}^{3}}{s}\{\left[s^2-{(M_{J/{\psi}}^2
-M_{\pi}^2)}^{2}\right]
\left[s^2-{(M_{D}^2-M_{\bar D}^2)}^{2}\right]\}^{\frac{1}{2}},\nn\\
&&\beta_{1}=\frac{2 [2\lambda_{Qq}(\lambda_{QQ}\eta^2+
\lambda_{qq}^{'}(1-\eta)^2)+\lambda_{qq}^{'}\lambda_{QQ}]}
{\lambda_{qq}^{'}+2\lambda_{Qq}+\lambda_{QQ}},\nn\\
&&\beta_{2}=\frac{2[2\lambda_{Qq}(\lambda_{QQ}^{'}\eta^2
+\lambda_{qq}(1-\eta)^2)+\lambda_{qq}\lambda_{QQ}^{'}]}
{\lambda_{QQ}^{'}+2\lambda_{Qq}+\lambda_{qq}},\nn\\
&&\Lambda_{1}=\frac{(x+\lambda_{qq})(\lambda_{qq}^{'}+2\lambda_{Qq}
+\lambda_{QQ})}{\lambda_{Qq}},\nn\\
&&\Lambda_{2}=\frac{(x+\lambda_{QQ})(\lambda_{QQ}^{'}+2\lambda_{Qq}
+\lambda_{qq})}{\lambda_{Qq}}.
\eea
In these:
\bea
&&\lambda_{Qq}=b_{D}^2/4=b_{\bar D}^2/4,\hspace{0.2cm}
\lambda_{qq}=b_{\pi}^2/4,\hspace{0.2cm}\lambda_{QQ}=b_{J/\psi}^2/4,
\hspace{0.2cm}\eta=\frac{m_{Q}}{m_{q}+m_{Q}},\nn\\
&&\lambda_{Qq}^{'}=\frac{x\lambda_{Qq}}{x+\lambda_{Qq}},
\hspace{0.2cm}\lambda_{qq}^{'}=\frac{x\lambda_{qq}}{x+\lambda_{qq}},
\hspace{0.2cm}\lambda_{QQ}^{'}=\frac{x\lambda_{QQ}}{x+\lambda_{QQ}}.
\eea

The cross-sections for the final channel are obtained from 
Eqs.~(\ref{scd})~and~(\ref{Sigma}). The total cross section for the reaction 
is a function of the center-of-mass energy and is obtained by summing over 
all possible final channels $\sigma_{tot}(s)=\sum_{f=1}^{4}\sigma_{fi}(s)$.
For numerical evaluations, the same parameter values as in 
Ref.\cite{qua} are used:
\bea
&&m_{Q}=1.67\hspace{0.2cm} GeV \hspace{0.2cm},\hspace{0.2cm}m_{q}=0.33
\hspace{0.2cm} GeV,\\
&&V_{0}\rightarrow \frac{3}{4}V_{0}=0.675\hspace{0.2cm} GeV,\hspace{0.2cm}
\hspace{0.2cm}C=0.24\hspace{0.2cm} GeV,\hspace{0.2cm}
\hspace{0.2cm}x=1.0\hspace{0.2cm} GeV^{-2},\\
&&\lambda_{QQ}=0.64\hspace{0.2cm} GeV^{-2},\hspace{0.2cm}
\hspace{0.2cm}\lambda_{qq}=2.4\hspace{0.2cm} GeV^{-2},\hspace{0.2cm}
\hspace{0.2cm}\lambda_{Qq}=1.7\hspace{0.2cm} GeV^{-2}.
\eea

In Fig.~1 the total cross sections for the $q\aq$ channel is shown as a 
function of the relative kinetic energy of the $J/\psi$ and the $\pi$ in the 
center-of-mass system, $E$. The dashed line refers to~Ref.~\cite{qua}, 
and the solid line is obtained with our post-prior symmetrical interaction. 

\begin{center}
\epsfxsize=10.0cm
\centerline{\epsfbox{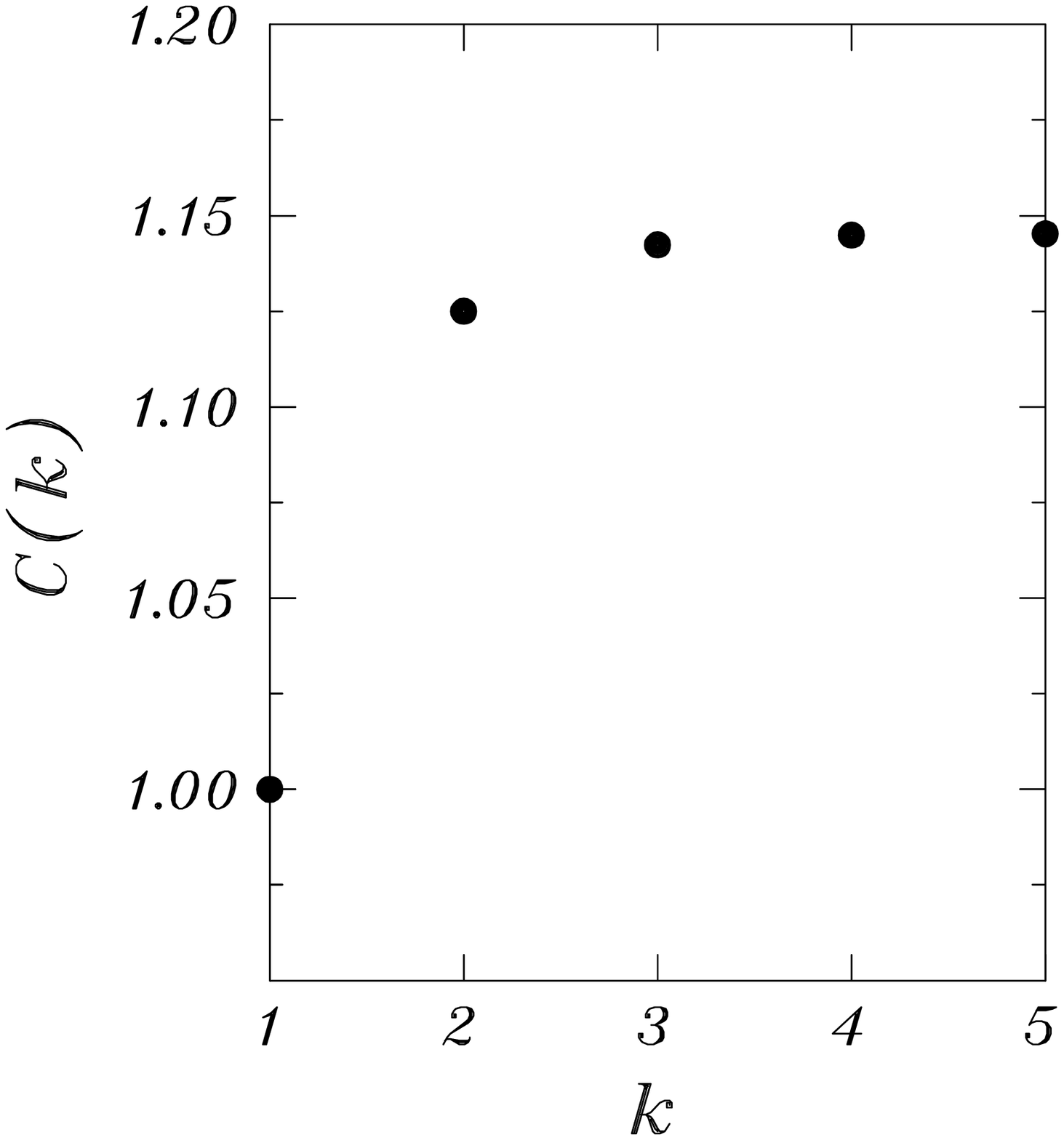}}
Figure 1. Total cross-sections for the prior and post-prior 
matrix elements.
\end{center}
\noindent

\vspace{0.5cm}
From this figure it is evident that the breaking of the post-prior 
symmetry might have a dramatic effect on the observables. We have 
explored the consequences of using the interaction also in the $qq$ and 
$\aq\aq$ channels~\cite{futureJPSI}, instead of using it only in the $q\aq$ 
channel as in Ref.~\cite{qua}.  We obtained the amusing result that the
lack of post-prior symmetry is to a large extent compensated by the inclusion 
of the $qq$ and $\aq\aq$ contributions in our explicitly post-prior 
symmetrical effective interaction, and the results are consistent with the 
values obtained in previous phenomenological 
calculations~\cite{qua}.

\newpage
\section{Mapping of baryons}
\label{sec:baryons}

As discussed in the previous sections, there are several appealing features of 
the Fock-Tani representation that make it particularly interesting for the 
treatment of composite hadronic interactions in quark models. The extension 
to the baryon case might be useful for several interesting 
applications in nuclear physics. The nucleon-nucleon interaction, for example,
exhibits a strongly repulsive short-distance core, which is attributed to the 
exchange of the $\omega$-meson. Since nucleons have a radii of about 0.8 fm 
and the range of the meson exchange force is $1/m_{\omega}\approx 0.2$ fm, it 
is natural to expect that the nucleon substructure will play a role at such 
short distances. Other examples include high density and/or temperature 
nuclear matter where the internal degrees of freedom of hadrons will
simultaneously be present with hadronic degrees of freedom.

We start with the definition of the single-composite baryon creation operator,
$B_{\alpha}^{\dagger}$, in terms of three quark creation operators
\bea
B^{\dagger}_{\al}=\frac{1}{\sqrt{3!}} 
\Psi_{\alpha}^{\mu_1\mu_2\mu_3}
q_{\mu_1}^{\dagger}q_{\mu_2}^{\dagger}q_{\mu_3}^{\dagger}.
\label{ft3}
\eea
$\Psi_{\alpha}^{\mu_1\mu_2\mu_3}$ is the baryon wave function, where the index
$\alpha$ identifies the quantum numbers of the baryon, and $\mu$ those of 
the quarks. The Fock-space amplitude is taken to be orthonormalized,
\bea
\lan  \alpha|\beta \ran= \Psi_{\alpha}^{*\mu_1\mu_2\mu_3 } 
\Psi_{\beta}^{\mu_1\mu_2\mu_3}=\delta_{\alpha \beta}\,.
\label{ft4_norm}
\eea
Using the quark anticommutation relations, Eq.~(\ref{qcom}), and the 
normalization condition above, it can easily be shown that the baryon 
operators satisfy the following noncanonical anticommutation relations
\bea
\{B_{\alpha},B_{\beta}^{\dagger}\}=\delta_{\alpha\beta}-\Delta_{\alpha\beta},
\hspace{1.5cm}
\{B_{\alpha},B_{\beta}\}&=&0,
\label{ft6}
\eea
where
\bea
\Delta_{\alpha\beta}=3\Psi_{\alpha}^{*\mu_1\mu_2\mu_3}\Psi_{\beta}^
{\mu_1\mu_2\nu_3}q_{\nu_3}^{\dagger}q_{\mu_3}-\frac{3}{2}\Psi_{\alpha}^
{*\mu_1\mu_2\mu_3}\Psi_{\beta}^{\mu_1\nu_2\nu_3}q_{\nu_3}^{\dagger}q_{\nu_2}^
{\dagger}q_{\mu_2}q_{\mu_3}. 
\label{ft7}
\eea
In addition, 
\bea
\{q_{\mu},B_{\alpha}^{\dagger}\} = \sqrt{ \frac{3}{2} }
\Psi_{\alpha}^{\mu\mu_2\mu_3}q_{\mu_2}^{\dagger}q_{\mu_3}^{\dagger},
\hspace{1.5cm}
\{q_{\mu},B_{\alpha}\}=0.
\label{ft8}
\eea
As in the case of mesons, the operator $\Delta_{\alpha \beta}$ is due to
the composite nature of the baryons. 

The unitary transformation is defined such that a single-baryon state 
$|\alpha \ran$ is transformed into an (``ideal") elementary-baryon state,
\bea
|\alpha\ran=B^{\dagger}_{\alpha}|0 \ran
\longrightarrow U^{-1}|\alpha \ran\equiv |\alpha)=
b^{\dagger}_{\alpha}|0) .
\label{single_bar}
\eea
The ideal baryon operators $b^{\dagger}_{\alpha}$ and $b_{\alpha}$,
by definition, satisfy the canonical anticommutation relations
\bea
\left\{b_{\alpha}, b^{\dagger}_{\beta}\right\}=\delta_{\alpha \beta},
\hspace{1.0cm}\left\{b_{\alpha}, b_{\beta}\right\}=0 .
\label{banticom}
\eea
The state $|0)$ is the vacuum of both $q$ and $b$ degrees of freedom in the new
representation. In addition,  the quark operators $q^{\dagger}$ and
$q$ are kinematically independent of the $b^{\dagger}_{\alpha}$ and 
$b_{\alpha}$ in the new representation
\bea
\{q_{\mu},b_{\alpha}\}=\{q_{\mu},b^{\dagger}_{\alpha}\}=0 .
\label{indep_bar}
\eea

The generator $F$ of the unitary transformation is again 
an anti-hermitian operator defined by
\bea
F=b^{\dagger}_{\alpha}O_{\alpha} - O_{\alpha}^{\dagger} b_{\alpha}.
\label{ft10}
\eea
The $O_{\al}$ operator is constructed in a iterative manner, order by order, 
in the baryon wave function by requiring that:
\bea
\{O_{\alpha}, O^{\dagger}_{\beta}\}=\delta_{\alpha\beta},\hspace{1.5cm}
\{O_{\alpha},O_{\beta}\}=\{O^{\dagger}_{\alpha},O^{\dagger}_{\beta}\}=0 .
\label{Oanti}
\eea
As in the case for mesons, this iterative procedure ensures that no secular 
terms appear. 
The explicit form of the operator $O_{\alpha}$ is obtained in a straightforward
calculation, following the same procedure as in the previous section.
Up to third order, it is given by
\bea
O_{\alpha}=B_{\alpha}+\frac{1}{2}\Delta _{\alpha \beta}B_{\beta}
-\frac{1}{2}B^{\dagger}_{\beta}[\Delta_{\beta\gamma},B_{\alpha}]B_{\gamma}.
\label{O_bar}
\eea
It is not difficult to verify that $O_{\al}$ has a canonical anticommutation
relation up to the third order in the $\Psi$'s. The next step is to transform 
the quark operators.

\subsection{Transformation of the quark operators}
\label{sec:operators_bar}

Given the generator $F$ of the baryon transformation to a certain order, 
one needs to evaluate
\bea
{\cal O}_{\rm FT}=U^{-1}\, {\cal O}\,U.
\label{eqmot_bar}
\eea
As discussed in the previous section for the mesons, because of 
Eq.~(\ref{Oanti}), the equations of motion for the operators $b_{\alpha}(t)$ 
and $O_{\alpha}(t)$ are manifestly symmetric,
\bea
 \frac{db_{\alpha}(t)}{dt} &=& [b_{\alpha}(t),F]= O_{\alpha}(t) ,
\hspace{1cm}
 \frac{dO_{\alpha}(t)}{dt}  = [O_{\alpha}(t),F]=-b_{\alpha}(t) ,
\label{ft14}
\eea
and can be trivially integrated, yielding
\bea
b_{\alpha}(t) &=& O_{\alpha} \sin t + b_{\alpha} \cos t,
\hspace{1cm}
O_{\alpha}(t) = O_{\alpha} \cos t - b_{\alpha} \sin t.
\label{ft17}
\eea
The transformation of the quark operators follows as in the previous section.
However, the transformed operators for the present case contain a few more 
terms than those for mesons. In order to simplify the presentation the results
are quoted for $t=-\pi/2$. 

The zeroth order operator is of course $q^{(0)}_\mu=q_\mu$. The first order 
one is
\beq
q^{(1)}_{\mu} = \delta_{\mu\mu_1}\,
\sqrt{\frac{3}{2}}\Psi^{\mu_{1}\mu_{2}\mu_{3}}_{\beta} 
q^{\dagger}_{\mu_{2}} q^{\dagger}_{\mu_{3}}
\left(b_{\beta} - B_{\beta}\right).
\label{f22}
\eeq
The solution to the second order equation leads to 
\bea
q^{(2)}_{\mu} &=& \delta_{\mu\nu_1}
\frac{3}{2}\Psi^{\ast\mu_{1}\mu_{2}\mu_{3}}_{\alpha}
\left[\Psi^{\nu_{1}\mu_{2}\mu_{3}}_{\beta}
(b^{\dagger}_{\alpha}q_{\mu_{1}} b_{\beta}
+ B^{\dagger}_{\alpha}q_{\mu_{1}}B_{\beta}
- 2 B^{\dagger}_{\alpha}q_{\mu_{1}}b_{\beta})\right.\nn\\
&-& \left.\Psi^{\nu_{1}\nu_{2}\mu_{3}}_{\beta}
\left(b^{\dagger}_{\alpha} q^{\dagger}_{\nu_{2}}q_{\mu_{2}}q_{\mu_{3}} 
b_{\beta} + B_{\alpha}^{\dagger }q_{\nu_2}^{\dagger}q_{\mu_2}q_{\mu_3}
B_\beta - 2B_\alpha^{\dagger}q_{\nu_2}^{\dagger}q_{\mu_2}q_{\mu_3}
b_\beta\right)\right]\,.
\eea
And finally, the solution of the third order equation is
\bea
&&q_\mu^{(3)} = \delta_{\mu\sigma_1}\sqrt{\frac{3}{2}}\Biggl\{
\frac{1}{2}\Psi_\alpha^{\sigma_1\mu_2\mu_3}q_{\mu_2}^{\dagger}
q_{\mu_3}^{\dagger}\Delta_{\alpha \gamma }
\left(b_\gamma-2B_\gamma\right)\nn\\
&&+\frac{3}{2}\Psi^{\ast\mu_1\mu_2\mu_3}_{\alpha}\Bigl[
\Psi^{\ast\nu_1\nu_2\nu_3}_{\beta}\Psi^{\sigma_1\mu_2\nu_3}_{\gamma}
(B_\alpha^{\dagger}-b_\alpha^{\dagger})
B_\beta^{\dagger}q_{\mu_1}q_{\mu_3}q_{\nu_2}q_{\nu_1}b_\gamma \nn\\
&&+\,\left(2\Psi_\beta^{\mu_1\mu_2\nu_3}\Psi_\gamma^{\sigma_1\nu_2\mu_3}
-\Psi_\beta^{\mu_1\nu_2\nu_3}\Psi_\gamma^{\sigma_1\mu_2\mu_3}\right)
\, q_{\nu _2}^{\dagger}q_{\nu_3}^{\dagger}
\left(-b_\alpha^{\dagger} b_\beta b_\gamma
+B_\alpha^{\dagger} B_\beta B_\gamma 
-B_\alpha^{\dagger} B_\beta b_\gamma
+b_\alpha^{\dagger} B_\beta b_\gamma \right) \nn\\  
&& + \, 2 \Psi_\beta^{\nu_1\mu_2\nu_3}
\Psi_\gamma ^{\sigma_1\nu_2\mu_3}q_{\nu_1}^{\dagger}q_{\nu_2}^{\dagger}
q_{\nu_3}^{\dagger}
\left(-b_\alpha^{\dagger} b_\beta b_\gamma 
+B_\alpha^{\dagger} B_\beta B_\gamma 
-B_\alpha^{\dagger} B_\beta b_\gamma    
+b_\alpha^{\dagger} B_\beta b_\gamma\right)q_{\mu_1} \Bigr] \Biggr\} .
\label{apena3}
\eea

This result concludes the evaluation of the transformed operators which
will be necessary to construct the Hamiltonian in the new
representation. This is done in the next section. 


\subsection{Effective baryon Hamiltonian}
\label{sec:HFT_bar}

The Fock-Tani Hamiltonian $H_{\rm FT}$ is obtained from the
quark-quark part of Hamiltonian in Eq.~(\ref{qHamilt}). The free-space 
eigenvalue equation for the single-baryon state is
\beq
H(\mu \nu;  \sig\rho )
\Psi _\alpha ^{ \sig\rho\lambda} =
\e_{[\alpha]} \Psi _{[\alpha]} ^{\mu \nu \lambda},
\label{ft24}
\eeq
where the Hamiltonian matrix $H(\mu \nu;  \sig\rho )$ is given by
\beq
H(\mu \nu , \sig\rho)=3\,[\,\delta _{[\mu] \sig }
\delta _{\nu\rho}\, T ([\mu]) + V_{qq}(\mu \nu;  \sig\rho)\,]. 
\label{ft25}
\eeq
$\e_{\alpha}$ is the total energy of the baryon. Here the notation is once 
again used so that there is no summation over repeated indices in square 
brackets.

The Fock-Tani Hamiltonian for baryons has a general structure similar to the 
$H_{\ft}$ of mesons, namely,
\bea
H_{\ft} =H_{q}+ H_{b} +H_{bq}.
\label{ft25_1}
\eea
The term $H_{q}$ involves only quark operators, $H_{b}$ only
ideal baryon operators, and $H_{bq}$ involves both quark and ideal baryon 
operators. 

In order to obtain the quark-quark interaction in the new representation, 
only the appropriate terms need be collected from the lowest order ones
$q^{\dagger\,(0)}_{\mu} q^{(1)}_{\mu}$ and
$q^{\dagger\,(1)}_{\mu} q^{(1)}_{\mu}$, from the transformation of
kinetic energy, and $q^{\dagger\,(1)}_{\mu}q^{\dagger\,(0)}_{\nu}
q^{(0)}_{\rho}  q^{(0)}_{\sigma}$ and $q^{\dagger\,(1)}_{\mu}  
q^{\dag\,(0)}_{\nu}q^{(0)}_{\rho}  q^{(1)}_{\sigma}$, from the transformation 
of the potential. This leads for the new quark-quark interaction the 
following expression
\bea
V_{qq}&=& \frac{1}{2} \,
V_{qq}(\mu\nu;\sig\rho)\,
q^{\dagger}_{\mu}q^{\dagger}_{\nu}q_{\rho}q_{\sig}
\nn\\
&&+
\frac{1}{6}\,\left[\frac{}{}
\Delta (\nu _1\nu _2\nu _3;\mu_1 \mu \nu )\,
H(\mu \nu;\mu_2 \mu_3)\,
+ H(\nu_1 \nu_2;\sig \rho)\,
\Delta (\sig \rho \nu_3;\mu _1\mu _2\mu _3)\,
\right. 
\nn\\
&&-
\left.\frac{}{}
\Delta (\nu _1\nu _2\nu _3;\tau\mu \nu )\,
H(\mu \nu;\sig \rho)\,
\Delta (\sig \rho \tau;\mu _1\mu _2\mu _3)\right]\,
q^{\dagger}_{\nu_{1}}q^{\dagger}_{\nu_{2}}q^{\dagger}_{\nu_{3}}
q_{\mu_{1}}q_{\mu_{2}}q_{\mu_{3}},
\label{ft25_2}
\eea
where $\Delta(\mu\nu\tau;\mu'\nu'\tau')$ is the bound state kernel for baryons,
\beq
\Delta(\mu \nu\tau;\mu'\nu'\tau')= \sum_{\alpha} \Psi^{\mu \nu\tau}_{\al}
\Psi^{\ast\mu' \nu'\tau'}_{\al} .
\label{ft25_3}
\eeq

\newpage
In the case that the $\Psi$'s are chosen to be the eigenstates of the 
microscopic quark Hamiltonian, Eq.~(\ref{ft25_2}) can be written in the 
following suggestive way
\bea
V_{qq}&=& \frac{1}{2} \,
V_{qq}(\mu\nu;\sig\rho)\,
q^{\dagger}_{\mu}q^{\dagger}_{\nu}q_{\rho}q_{\sig}
- \sum_{\alpha} {\cal E}_{\alpha} B^{\dagger}_{\alpha} B_{\alpha}.
\label{ft25_4}
\eea
As in the case of mesons, it is not difficult to show that the spectrum of the
modified quark Hamiltonian $H_{q}$ in Eq.~(\ref{ft25_1}) is positive 
semidefinite. 

Note that because $V_{qq}(\mu\nu;\sig\rho)$ is a two-body interaction, the 
creation and annihilation operators with sub-index~3 can always be contracted 
with each other in Eq.~(\ref{ft25_2}), such that the new quark-quark 
interaction, $V_{qq}$, remains in a two-body form . 

Among the various baryon-quark Hamiltonians $H_{bq}$ in Eq.~(\ref{ft25_1}), 
the two interesting ones are the single-baryon break-up and the binary 
collision 
break-up. By collecting the appropriate lowest order terms of the 
transformation of the kinetic and potential terms, the 
single-baryon break-up interaction (single-break) is obtained as
\bea
H_{\rm single-break}&=&\frac{1}{\sqrt{6}} \left[
H(\mu_1\mu_2;\sigma\rho) \Psi^{\sigma\rho\mu_3}_{\beta}\right. \nn\\
&-& \left. H(\mu\nu;\sigma\rho)\Psi^{\sigma\rho\tau_3}_{\beta}
\Delta(\mu_1\mu_2\mu_3;\mu\nu\tau_3)\right]
q^{\dagger}_{\mu_1}q^{\dagger}_{\mu_2}q^{\dagger}_{\mu_3}b_{\beta}.
\label{b-up}
\eea
The hermitian conjugate of $H_{\rm single-break}$ represents a baryon 
recombination. When $\Psi$ is chosen to be a stationary state of the 
microscopic quark Hamiltonian, Eq.~(\ref{b-up}) becomes
\beq
H_{\rm single-break} = 0.
\label{b-up:0}
\eeq
This result reflects the stability of the baryon state to spontaneous decay
in the absence of external perturbations. For a baryon in the 
environment produced in a heavy-ion collision, for example, one would be 
interested in the modifications on the free-space wave function properties of 
the nucleons, and obviously the effective Hamiltonian of Eq.~(\ref{b-up}) 
would be relevant in such a situation.

In the same way, a binary baryon collision break-up (binary-break) term can be
found,
\bea
V_{\rm binary-break} &=& \frac{1}{2} V_{\qq}(\mu\nu;\sigma\rho)\,
q^{\dagger\,(0)}_{\mu}  q^{\dagger\,(0)}_{\nu}
  q^{(1)}_{\rho}  q^{(1)}_{\sigma} \nn\\
&=& \frac{3}{4}\Psi_{\alpha}^{\rho   \nu _2\mu _3}
\Psi_{\beta}^{\sigma\nu _2\nu _3} V_{qq}(\mu\nu;\sigma\rho)\,
q^{\dagger}_{\mu}  q^{\dagger}_{\nu} q^{\dagger}_{\mu_2} 
q^{\dagger}_{\mu_3}  q^{\dagger}_{\nu_2} q^{\dagger}_{\nu_3}
b_{\alpha} b_{\beta}\,.
\label{bin-coll}
\eea
In a hot and/or dense medium, where constituents and composites can be 
simultaneously present in the system break-up terms may play an important role.

The piece of the effective Hamiltonian that contains only ideal baryon 
operators, represented by the second term of Eq.~(\ref{ft25_1}), can be 
written as:
\bea
H_{b}=H^{(0)}_{b} + V_{bb}.
\label{ft25.b}
\eea
$H^{(0)}_{bb}$ is the ``non-interacting" term, obtained by  substituting the 
transformed operators into the expression
\beq
T(\mu) q_\mu ^{\dagger (1)}(t)q_\mu ^{(1)}(t)
+\frac 12
V_{qq}(\mu \nu;  \sig\rho )
q_\mu^{\dagger (1)}(t)q_\nu ^{\dagger (0)}(t)
q_\rho ^{(0)}(t)q_\sig ^{(1)}(t).
\label{ft25.c}
\eeq
After putting the expression in normal order, a term is obtained that 
contains only ideal baryon operators,
\beq
H^{(0)}_{b} = H^{(0)}_{\ft}(\alpha;\beta)\,b^{\dagger}_{\alpha}b_{\beta},
\label{ft25.d}
\eeq
where
\beq
H^{(0)}_{\ft}(\alpha;\beta)=\Psi^{\ast\mu\nu\tau}_{\al} 
H(\mu\nu; \sig\rho)\Psi^{\sig\rho\tau}_{\beta} .
\label{ft25.e}
\eeq

The baryon-baryon potential is obtained from the expression 
\bea
&&T(\mu)\left[q_\mu^{\dagger (3)}(t)q_\mu^{(1)}(t)+q_\mu^{\dagger (1)}(t)
q_\mu^{(3)}(t)\right] 
+\frac 12 V_{qq}(\mu \nu;\sig\rho)
\left[q_\mu^{\dagger (1)}(t)q_\nu^{\dagger (1)}(t)q_\rho^{(1)}(t)
q_\sig^{(1)}(t) \right.\nn\\  
&&+q_\mu^{\dagger (1)}(t)q_\nu^{\dagger (0)}(t)q_\rho^{(2)}(t)
q_\sig ^{(1)}(t)
+q_\mu^{\dagger (1)}(t)q_\nu^{\dagger (2)}(t)q_\rho^{(0)}(t)
q_\sig^{(1)}(t)\nn\\  
&&+q_\mu ^{\dagger (3)}(t)q_\nu^{\dagger(0)}(t)q_\rho^{(0)}(t)
q_\sig ^{(1)}(t)\left.
+q_\mu ^{\dagger (1)}(t)q_\nu^{\dagger (0)}(t)q_\rho^{(0)}(t)q_\sig^{(3)}(t)
\right].
\label{ft26}
\eea
Substituting the transformed quark operators, and putting the resulting
expression in normal order, terms are obtained that again involve only ideal 
baryon operators. The total baryon Hamiltonian may be written in the form
\beq
H_b=\Psi^{*\mu\nu\lambda}_{\alpha}H(\mu\nu;\sigma\rho)
\Psi^{\sigma\rho\lambda}_{\beta}\,b^{\dagger}_{\alpha}b_{\beta}+
\frac{1}{2}\,V_{bb}(\al\beta;\delta\gamma)\,b^{\dagger}_{\al}
b^{\dagger}_{\beta}b_{\gamma}b_{\delta},
\label{Hb}
\eeq
where $V_{bb} \equiv V^{dir}_{bb}+V^{exc}_{bb}+V^{int}_{bb}$ is the 
effective baryon-baryon potential, which is divided we divide again into  
direct, exchange, and intra-exchange parts, where
\newpage
\bea
V^{dir}_{bb}(\alpha\beta;\delta\gamma)&=& 9 \, V_{qq}(\mu \nu; \sigma \rho)
\Psi^{\ast\mu\mu_{2}\mu_{3}}_{\alpha}\Psi^{\ast\nu\nu_{2}\nu_{3}}_{\beta}
\Psi^{\rho\nu_{2}\nu_{3}}_{\gamma}\Psi^{\sigma\mu_{2}\mu_{3}}_{\delta},\\
\label{Vdir_bar}
V^{exc}_{bb}(\alpha\beta;\delta\gamma) &=&
9 \, V_{qq}(\mu \nu; \sigma \rho)
\left[2\,
\Psi^{\ast\mu\nu\mu_{3}}_{\alpha}\Psi^{\ast\nu_{1}\nu_2\nu_{3}}_{\beta} 
\Psi^{\rho\nu_2\nu_3}_{\gamma}\Psi^{\nu_1\sigma\mu_3}_{\delta}
\right.\nn\\
&&
-\Psi^{\ast\mu\mu_{2}\mu_{3}}_{\alpha}\Psi^{\ast\nu\nu_{2}\nu_{3}}_{\beta}
\left( \Psi^{\sigma\nu_{2}\nu_{3}}_{\gamma}\Psi^{\rho\mu_{2}\mu_{3}}_{\delta}
+4\Psi^{\rho\nu_{2}\mu_{3}}_{\gamma}\Psi^{\sigma\mu_{2}\nu_{3}}_{\delta}
\right)
\nn\\
&& \left.
-2 \,
\Psi^{\ast\mu\mu_{2}\mu_{3}}_{\alpha}\Psi^{\ast\nu_{1}\nu_2\nu}_{\beta} 
\Psi^{\nu_{1}\nu_2\mu_{3}}_{\gamma}
\Psi^{\sigma\mu_2\rho}_{\delta}
\right],
\label{Vexch_bar}
\\
V^{int}_{bb}(\alpha\beta; \delta\gamma)&=&-6 \, H(\mu \nu ;\sigma\rho)
\left(\Psi_{\alpha}^{*\mu\nu\mu_3}\Psi_{\beta}^{*\nu_1\nu_2\nu_3}
\Psi_{\gamma}^{\nu_1\nu_2\mu_3}\Psi_{\delta}^{\sigma\rho\nu_3}\right. \nn\\
&&\,\,\,\,+\left. \Psi_{\alpha}^{*\mu\mu_2\nu}\Psi_{\beta}^{*\nu_1\nu_2\nu_3}
\Psi_{\gamma}^{\nu_1\nu_2\rho}\Psi_{\delta}^{\sigma\mu_2\nu_3}
+\Psi_{\alpha}^{*\mu\mu_2\mu_3}\Psi_{\beta}^{*\nu_1\nu_2\nu}
\Psi_{\gamma}^{\nu_1\nu_2\mu_3}\Psi_{\delta}^{\sigma\mu_2\rho}\right).
\label{Vintr}
\eea

Similar to the meson case, it can be shown that if the $\Psi$'s are chosen to 
be the eigenstates of the microscopic quark Hamiltonian, the intra-exchange 
term $V_{bb}^{int}$ is precisely canceled by orthogonality corrections at 
lowest order.

In the next subsection, our nucleon-nucleon result is compared with the
one obtained in Ref.~\cite{QBD_bar}. In the quark model used in 
Ref.~\cite{QBD_bar}, the $\Psi$'s are taken to be nonrelativistic s-wave 
gaussians, and the microscopic quark-quark interaction is the spin-spin part 
of the nonrelativistic reduction of the one-gluon exchange. 

\subsection{An effective nucleon-nucleon potential}
\label{sec:effnn}

According to our notation, the nucleon wave function used in 
Ref.~\cite{QBD_bar} is given by
\bea
\Psi_\alpha ^{\mu _1\mu _2\mu _3} \equiv 
\delta ({\bp}_\alpha -{\bp}_1-{\bp}_2-{\bp}_3)\,
{\caln ({\bp}_\alpha )} \frac{\epsilon^{c_{\mu_1}c_{\mu_2}c_{\mu_3}}} 
{\sqrt{3!}} \,
\epsilon^{c_{\mu _1}c_{\mu _2}c_{\mu _3}}
\,\chi_\alpha ^{I_{\mu_1}I_{\mu _2}I_{\mu _3}}\,
\phi ({\bp}_1)\phi({\bp}_2)\phi({\bp}_3),
\label{qbdf5}
\eea
where $\eps^{c_{\mu _1}c_{\mu _2}c_{\mu _3}}$ is the color 
antisymmetric  tensor; $\chi_\alpha ^{I_{\mu_1}I_{\mu _2}I_{\mu _3}}$ 
is the Clebsch-Gordan coefficient of spin-isospin, the $\phi$'s are the 
single-quark wave functions, and $\caln({\bp}_\alpha )$ is a 
normalization function.  The explicit form of the momentum-space 
single-quark functions is
\beq
\phi ({\bp})=\left(\frac{b^2}{\pi}\right)^{3/4} \,
\exp \left( -\frac{b^2 \bp^2}{2} \right).
\label{qbdf6}
\eeq
With such a form, $\caln({\bp})$ is obtained to be
\beq
\caln({\bp})=\left(\frac{3\pi}{ b^{2}}\right)^{3/4}\,
\exp \left( \frac{b^2 \bp^2}{6}\right).
\label{qbdf6.1}
\eeq
The constant $b$ is related to the mean-square radius of the baryon by
$b^2= \lan r^2 \ran $.

Using a local quark-quark interaction $V_{qq}$ as in Ref.~\cite{QBD_bar}, i.e.
an interaction that depends only on the momentum transfer, the 12-dimensional 
integral over the quark coordinates in the expression of the effective 
potential can be  evaluated analytically, because the single-quark wave 
functions are gaussians. Then the NN potential can be written in the form
\bea
V_{NN}&=&\frac{1}{2}\int d{\bp}_1\,\cdots d{\bp}_4\,
\delta({\bp}_1 + {\bp}_2 - {\bp}_3 - {\bp}_4)\,
\lan  \lambda_3\lambda_4|V_{NN}({\bsigma}, {\btau}, {\bp}_1\cdots
{\bp}_4)|\lambda_1\lambda_2 \ran
\nn\\&\times & 
b^{\dagger}_{\lambda_4}({\bp}_4)\,b^{\dagger}_{\lambda_3}({\bp}_3)
\,b_{\lambda_2}({\bp}_2)\,b_{\lambda_1}({\bp}_1),
\label{VNNNL}
\eea
where $\lambda=(M_S,M_T)$ and
\bea
V_{NN}(\bsigma, \btau, {\bp}_1\cdots{\bp}_4)=
\sum_{i=1}^{5}\, \omega_{i}(\bsigma,\btau)
v_{i}({\bp}_1\cdots{\bp}_4).
\label{VNNNL_2}
\eea
The operators $\omega_i({\bsigma},{\btau})$ are obtained from the
sum over the quark color-spin-flavor indices; $\bsigma_N$ and $\btau_N$ are 
nucleon spin and isospin operators. The sum over the quark spin-isospin
quantum numbers can be evaluated in closed form by making use of the elegant 
technique of Ref.~\cite{hol}. The spatial functions can be written as
\bea
&&v_1({\bp}_1\cdots{\bp}_4) = F({\bq}^2)\,
V_{\qq}({\bq})\,F({\bq}^2),\\
&&v_i({\bp}_1\cdots{\bp}_4) = \left(\frac{3b^2}{4\pi}\right)^{3/2}\,
N({\bp}_1\cdots{\bp}_4) \, I_i({\bp}_1\cdots{\bp}_4),
\eea
where ${\bq}={\bp}_3-{\bp}_1={\bp}_2-{\bp}_4$ is the momentum transfer, 
$F({\bq}^2)$ is the nucleon form factor,
\bea
F({\bq}^2)=\exp\left(-\frac{b^2}{6}{\bq}^2\right),
\label{FormFac}
\eea
and $N({\bp}_1\cdots{\bp}_4)$ is given by
\bea
N({\bp}_1\cdots{\bp}_4)=\exp\left[-b^2\left(\frac{1}{2}{\bp}_4^2 + 
\frac{1}{4}{\bp}_3^2 +\frac{7}{12}{\bp}_2^2 - 
\frac{1}{6}{\bp}_1^2-{\bp}_4{\bcdot}{\bp}_2
-\frac{1}{3}{\bp}_3{\bcdot}{\bp}_2+\frac{1}{3}{\bp}_4{\bcdot}{\bp}_3\right)
\right]\,,
\label{Np1p4}
\eea
and $I_i({\bp}_1\cdots{\bp}_4), i=2,\cdots ,5$ are the integrals
\bea
I_{2}({\bp}_1\cdots{\bp}_4)&=&\int\,d\bold q \,V_{\qq} (\bq)\, 
\exp\left[-b^{2} \left(q^2 - \bq\bcdot(\bp_1-\bp_2)\right)\right]\,,\\
I_{3}(\bp_1\cdots\bp_4)&=&\int\,d\bq \,V_{\qq} (\bq)\, 
\exp\left[-b^2\left(\frac{3}{4} \bq^2 + \frac{1}{2}
\bq\bcdot(\bp_{3}-\bp_{2})\right)\right]\,,\\
I_{4}(\bp_{4}\cdots\bp_{4})&=&\int\,d\bq \,V_{\qq} (\bq)\, 
\exp\left[-b^2\left(\frac{11}{16} \bq^2 +
\bq\bcdot(\txs\frac{1}{2} \bp_{4}+\frac{1}{4}\bp_{3}-
\frac{3}{4}\bp_{2})\right)\right]\,,\\
I_{5}(\bp_{1}\cdots\bp_{4})&=&\int\,d\bq \,V_{\qq} (\bq)\, 
\exp\left[-b^2\left(\frac{11}{16} \bq^2 -
\bq\bcdot(\txs\frac{1}{2} \bp_{4}-\frac{1}{4}\bp_{3}
-\frac{1}{4}\bp_{2})\right)\right]\,.
\label{aplic_49}
\eea

When using $V_{qq}({\bq})=V_0=\text{constant}$, which represents a contact 
interaction, such as the spin-spin term from the one-gluon exchange used
in Ref.~\cite{QBD_bar}, the following expressions for the $v_i$'s
are obtained
\bea
v_{1}(\bp,\bp^{\prime})&=&v_{3}(\bp,\bp^{\prime})=
\,V_0\,\exp\left[-\frac{b^2}{3}(\bp-\bp^{\prime})^2\right]\,,
\nn\\
v_{2}(\bp,\bp^{\prime})&=&\,V_0\,
\left(\frac{3}{4}\right)^{3/2}\exp\left[
-\frac{b^2}{6}(\bp^{2} + \bp^{\prime\,2} )\right]\,,
\nn\\
v_{4}(\bp,\bp^{\prime})&=&v_{5}(\bp^{\prime},\bp)=\,V_0\,
\left(\frac{12}{11}\right)^{3/2}\exp\left[
-\frac{2b^2}{11}(\bp-\bp^{\prime})^2
-\frac{b^2}{33}(\bp^2 +7\bp^{\prime\,2})
\right].
\label{aplic_55}
\eea

Now, when the color-spin-flavor dependence of the interaction is 
equal to the spin-spin term of the one-gluon exchange, 
Eq.~(\ref{mic3}), the spin-isospin functions $\omega_{i}(\bsigma,\btau)$ are 
given by
\bea
\omega_1 &=&0, \nn\\
\omega_{2}&=&\frac{1}{12}\left[\left(1+\frac{1}{9}{\btau}^{1}_{N}\bcdot
{\btau}^{2}_{N}\right)+ \frac{1}{3}\left(1+ \frac{1}{9}{\btau}^{1}_{N}\bcdot
{\btau}^{2}_{N}\right){\bsigma}^{1}_{N}\bcdot
{\bsigma}^{2}_{N}\right],\nn\\
\omega_{3}&=&\frac{3}{4}\left[\left(1+\frac{1}{9}{\btau}^{1}_{N}\bcdot
{\btau}^{2}_{N}\right)- \frac{1}{27}\left(1+ \frac{25}{9}{\btau}^{1}_{N}
\bcdot{\btau}^{2}_{N}\right){\bsigma}^{1}_{N}\bcdot{\bsigma}^{2}_{N}\right],
\nn\\
\omega_{4}&=&\omega_{5}=
\frac{1}{4}\left[\left(1-\frac{1}{9}{\btau}^{1}_{N}\bcdot
{\btau}^{2}_{N}\right)-\frac{1}{9}\left(1- \frac{5}{9}{\btau}^{1}_{N}\bcdot
{\btau}^{2}_{N}\right){\bsigma}^{1}_{N}\bcdot{\bsigma}^{2}_{N}\right],
\label{gra4}
\eea
Note that $\omega_{1}=0$ because there is no one-gluon exchange between 
colorless baryons.

On-shell, i.e., when ${\bp}^2={\bp'}^2$, this result is precisely the same  
as obtained within the quark Born diagram method for the 
T-matrix~\cite{QBD_bar}. For the off-shell case, it is
evident that our interaction is post-prior symmetric. The symmetry is of 
importance for calculations beyond the Born approximation, where the 
potential is iterated in a Lippmann-Schwinger equation.

In the next section orthogonality corrections to the lowest order hadron-hadron
Hamiltonians will be derived. The study of the hadron-hadron interaction 
using the constituent quark model has been traditionally been done with
the resonating group method (RGM). There is an extensive 
literature on the subject; two good review articles are given in 
Ref.~\cite{clust}. The RGM will be used in the next section to make contact 
with the Fock-Tani representation, and to illustrate in a transparent way the 
physical meaning of the orthogonality corrections for the effective 
meson-meson interaction.


\newpage 
\section{Orthogonality corrections}
\label{sec:ortho}

Orthogonality corrections appear in the form of terms proportional to the 
bound state kernels $\Delta(\mu\nu;\sigma\rho)$ and 
$\Delta(\mu \nu\lambda;\sigma\rho\tau)$, and have the effect, among others, 
of weakening the ``intra-exchange'' inter\-actions. An example of this effect 
is the renormalization of the microscopic interaction, shown in 
Eqs.~(\ref{modqaq})~and~(\ref{ft25_2}). In the hadron-hadron interaction, they
reflect the Pauli principle among the constituents in the
different clusters. In this section, the lowest-order corrections (in an 
expansion in powers of the bound state kernels $\Delta$) for the effective 
meson and baryon Hamiltonians are obtained, and the RGM is used to evaluate 
the magnitude of higher oder terms. 

Before entering into the derivation of the orthogonality corrections within the
Fock-Tani representation, use is made of the RGM for the meson-meson 
scattering. The RGM is extensively used in the context of hadron-hadron 
scattering~\cite{clust}, and the meaning and origin of these corrections 
is particularly transparent within this method. Of course, as will become 
clear at the end of the discussion, orthogonality corrections apply to all 
pieces of the effective Hamiltonian, not only to the effective hadron-hadron 
interaction, and can be systematically derived from within the Fock-Tani 
representation.

In a RGM calculation the two-cluster state is introduced by writing 
\beq
|\Lambda \ran=\frac{1}{\sqrt{2}}\psi^{\alpha\beta}_{\Lambda}
M^{\dagger}_{\alpha}M^{\dagger}_{\beta}|0 \ran,
\label{RGManz}
\eeq
where $\psi^{\alpha\beta}_{\Lambda}$ is the ansatz wave function for the 
meson pair; it describes the c.m. and relative motions of the two 
meson clusters. The $M^{\dagger}$'s are the meson creation operators as 
defined in Eq.~(\ref{Mop}). $\Lambda$ identifies the set of quantum numbers 
of the two-cluster state. Using the commutation relation of the meson 
operators, Eq.~(\ref{Mcom}), the normalization condition for the
$\psi^{\alpha\beta}_{\Lambda}$ is obtained to be
\beq
\lan  \Lambda|\Lambda' \ran=\psi^{*\alpha\beta}_{\Lambda}N(\alpha\beta; 
\alpha'\beta')\psi^{\alpha'\beta'}_{\Lambda'}=\delta_{\Lambda'\Lambda},
\label{NormPsi}
\eeq
where $N(\alpha\beta ; \alpha'\beta')$ is the ``normalization kernel", given
by
\beq
N(\alpha\beta ; \alpha'\beta')=\delta_{\alpha\alpha'}\delta_{\beta\beta'}
-N_E(\alpha\beta; \alpha'\beta')= \delta_{\alpha\alpha'}\delta_{\beta\beta'}
-\Phi^{*\mu\nu}_{\alpha}\Phi^{*\rho\sigma}_{\beta}\Phi^{\mu\sigma}_{\beta'}
\Phi^{\rho\nu}_{\alpha'}.
\label{NormKern}
\eeq
The exchange kernel $N_E(\alpha\beta; \alpha'\beta')$ comes from the 
noncanonical part of the meson commutation relation of Eq.~(\ref{Mcom}), and
it reflects the Pauli principle among the quarks and antiquarks in the
clusters $\alpha$ and $\beta$. The equation of motion for 
$\psi^{\alpha\beta}_{\Lambda}$ is determined by means of the variational 
principle
\beq
\delta\lan  \Lambda|(H-E_{\Lambda})|\Lambda \ran=0,
\eeq
which leads to the RGM equation,
\beq
\left[H_{RGM}(\alpha\beta; \gamma\delta)-E_{\Lambda}
N(\alpha\beta; \gamma\delta)\right]\psi^{\gamma\delta}_{\Lambda}=0,
\label{eqmrg}
\eeq
with
\beq
H_{RGM}(\alpha\beta; \gamma\delta)= T_{RGM}(\alpha\beta; \gamma\delta)
+V_{mm}(\alpha \beta;\gamma \delta),
\eeq
where the  kinetic term $T_{RGM}(\alpha\beta; \gamma\delta)$ is given by
\beq
T_{RGM}(\alpha\beta; \gamma\delta)= \delta_{\beta\delta}
\Phi_{\alpha}^{*\mu\nu}H(\mu\nu; \mu'\nu')
\Phi_{\gamma}^{\mu'\nu'}+\delta_{\alpha\gamma}
\Phi_{\beta}^{*\mu\nu}H(\mu\nu; \mu'\nu')
\Phi_{\delta}^{\mu'\nu'},
\label{TRGM}
\eeq
and the potential terms $V_{mm}(\alpha \beta;\gamma \delta)$ are precisely
equal to the Fock-Tani potentials given in Eqs.~(\ref{Vmm})-(\ref{Vintra}).

The two-meson wave function is not normalized in the usual quantum 
mechanical way, because of the presence of normalization kernel in
Eq.~(\ref{NormPsi}). It is common practice~\cite{clust} to introduce a 
``renormalized" wave function defined as
\bea
\bar{\psi}^{\alpha\beta}_{\Lambda} \equiv 
N^{\frac{1}{2}}(\alpha\beta; \alpha'\beta')\psi^{\alpha'\beta'}_{\Lambda},
\eea
where $N^{1/2}$ is the square root of the RGM normalization kernel. This 
clearly leads to
\beq
\bar{\psi}^{*\alpha\beta}_{\Lambda'}\bar{\psi}^{\alpha\beta}_{\Lambda}=
\delta_{\Lambda'\Lambda}.
\eeq
In terms of the renormalized wave function, the RGM equation can be rewritten 
as
\beq
\left[\bar{H}_{RGM}(\alpha\beta; \gamma\delta)-E_{\Lambda}
\delta_{\alpha\gamma}\delta_{\beta\delta}\right]
\bar{\psi}^{\gamma\delta}_{\Lambda}=0,
\label{RenoRGMEq}
\eeq
where the  ``renormalized" RGM Hamiltonian is defined as
\beq
\bar{H}_{RGM}(\alpha\beta; \gamma\delta) \equiv N^{-\frac{1}{2}}(\alpha
\beta; \alpha'\beta')H_{RGM}(\alpha'\beta'; \gamma'\delta')
N^{-\frac{1}{2}}(\gamma'\delta'; \gamma\delta).
\label{RenoRGMH}
\eeq
Now, if $N^{-\frac{1}{2}}$ is expanded in Eq.~(\ref{RenoRGMH}) according 
to
\beq
N^{-\frac{1}{2}}=\left(1-N_E\right)^{-\frac{1}{2}}=1+\frac{1}{2}
N_E+\frac{3}{8}N_E^2 + \cdots,
\label{expN-1/2}
\eeq
where $N_E$ is the exchange kernel defined in Eq.~(\ref{NormKern}), and 
only the first order term is retained, the lowest order correction to 
the RGM Hamiltonian is given by
\bea
&&\bar{H}_{RGM}(\alpha\beta; \gamma\delta)= T_{RGM}(\alpha\beta; \gamma\delta)
+V^{dir}_{mm}(\alpha \beta;\gamma \delta)
+ V^{exc}_{mm}(\alpha \beta;\gamma \delta) \nn\\
&&\hspace{0.6cm}-\frac{1}{2}\left\{\Phi_{\alpha}^{* \mu \nu}
\Phi_{\beta}^{* \rho \sigma}\left[H(\mu \nu;\mu'\nu')
-H(\mu\nu;\lambda\tau)\Delta(\lambda\tau;\mu^{\prime}\nu^{\prime})\right]
\Phi_{\delta}^{\mu' \sigma}\Phi_{\gamma}^{\rho \nu'}
+(\alpha\leftrightarrow\beta;\gamma\leftrightarrow\delta)\right\}\nn\\
&&\hspace{0.6cm}-\frac{1}{2}\left\{\Phi_{\alpha}^{* \mu \sigma}
\Phi_{\beta}^{* \rho \nu}\left[H(\mu \nu ;\mu' \nu')
-\Delta(\mu\nu;\lambda\tau)H(\lambda\tau;\mu^{\prime}
\nu^{\prime})\right]\Phi_{\delta}^{\mu' \nu'}
\Phi_{\gamma}^{\rho \sigma} 
+(\alpha\leftrightarrow\beta;\gamma\leftrightarrow\delta)\right\}\,.
\eea
If the  $\Phi$'s are chosen to be the eigenstates of the microscopic quark 
Hamiltonian, the intra-exchange term $V_{mm}^{int}$ is cancelled (see
Eq.~(\ref{Vintra}). This cancelation is the main effect of the renormalization
of the wave function, higher order terms in the expansion give small 
corrections. This is explicitly demonstrated in the following two examples.

The derivation within the Fock-Tani representation of the corrections 
discussed above is trivial, since these always appear in such a form that the 
microscopic Hamiltonian acts on a bound state kernel. With a little of 
experience with the manipulation of the equations of motion of the quark 
operators, the relevant terms in these equations can easily be identified. 
It is easy to convince oneself that the lowest order corrections for the 
effective meson-meson potential come from terms in the microscopic 
quark-antiquark interaction of the form 
$q^{(1)\dagger}_{\mu}\aq^{(0)\dagger}_{\nu}\aq^{(0)}_{\rho}q^{(5)}_{\sigma} 
+ \text{h.c.}$. In order to obtain 
$q^{(5)}$ and $q^{(5)\dagger}$, the generator $F$ of the
transformation at fourth order is needed. 

It is not difficult to show that the fourth order $O$ operator for mesons is 
given by
\bea
O^{(4)}_{\alpha}=\frac{3}{8}\,\Delta_{\alpha\beta}\Delta_{\beta\gamma}
M_{\gamma}
-\frac{1}{8}\,M^{\dagger}_{\beta}[\Delta_{\alpha\gamma},\Delta_{\beta\delta}]
M_{\delta}M_{\gamma}
+\frac{1}{4}\,M^{\dagger}_{\beta}[M_{\alpha},
[\Delta_{\beta\gamma},M^{\dagger}_{\delta}]]M_{\gamma}M_{\delta}\,,
\label{mes4th}
\eea
In the proof of the commutation relation of Eq.~(\ref{comO}) up to fourth 
order, it is useful to make use of the Jacobi identity for bosonic operators 
$A,B,$ and $C$,
\beq
[A,\,[B,C]\,]+[C,\,[A,B]\,]+[B,\,[C,A]\,]=0,
\eeq

Next, the fifth order quark equation of motion is obtained, and only the 
terms that are relevant for the lowest order orthogonality corrections are
retained. These come with an antiquark creation operator $\aq^{\dagger}$ and 
three ideal meson operators. This is because the $\aq^{(0)}_{\rho}=\aq_{\rho}$
must be canceled in the expression 
$q^{(1)\dagger}_{\mu}\aq^{(0)\dagger}_{\nu}\aq^{(0)}_{\rho}q^{(5)}_{\sigma}$, 
and since $q^{(1)\dagger}_{\mu}\aq^{(0)\dagger}_{\nu} \sim m^{\dagger}$, extra
three ideal meson operators are needed to form an effective meson-meson 
interaction.

The equation of motion for the quark operator to fifth order, retaining only 
the relevant terms for the orthogonality corrections, is given by
\bea
{dq^{(5)}_{\mu}(t) \over dt}\Bigg|_{\text{relev}} &=& 
-\delta_{\mu\mu'1}\frac{1}{2} \Bigl[
M^{(0)}_{\beta}(t)\Phi^{\mu'\nu}_{\alpha}\aq^{\dagger(0)}_{\nu}(t)
\Delta^{(4)}_{\alpha\beta}(t)m^{(0)}_{\beta}(t)\nn\\
&+& \Delta(\mu'\nu,\rho'\sigma')\Phi^{\ast\rho\sigma}_{\alpha}
\Phi^{\rho'\sigma}_{\beta}\Phi^{\rho\sigma'}_{\gamma}
\aq^{(0)\dagger}_{\nu}(t)M^{(0)\dagger}_{\alpha}(t)M^{(0)}_{\beta}
m^{(0)}_{\gamma}(t)
\Bigr].
\label{q5th}
\eea
In the same way, a similar equation is obtained for $\aq^{(5)}_{\mu}(t)$, 
which is necessary for the transformation of the kinetic energy operator. 
Integrating the equations and  taking $t=-\pi/2$, results in 
\bea
&&T(\mu)q^{\dagger(5)}_{\mu}q^{(1)}_{\mu}+T(\nu)\aq^{\dagger(5)}_{\nu}
\aq^{(1)}_{\nu}+ \left[V_{q\aq}(\mu\nu;\rho\sigma)q^{(5)\dagger}_{\mu}
\aq^{(0)\dagger}_{\nu}\aq^{(0)}_{\sigma}q^{(5)}_{\rho}+\text{h.c.}\right] \nn\\
&&=+\frac{1}{2}\left\{
\left[\Phi_{\alpha}^{*\mu\sigma}
\Phi_{\beta}^{*\rho\nu}\Delta(\mu\nu;\lambda\tau)H(\lambda\tau;\mu^{\prime}
\nu^{\prime})\Phi_{\delta}^{\mu^{\prime}\nu^{\prime}}
\Phi_{\gamma}^{\rho\sigma} + 
(\alpha\leftrightarrow\beta;\gamma\leftrightarrow\delta)\right]\right.\nn\\
&&+\left.\left[\Phi_{\alpha}^{* \mu \nu}\Phi_{\beta}^{*\rho\sigma}
H(\mu\nu;\lambda\tau)\Delta(\lambda\tau;\mu^{\prime}\nu^{\prime})
\Phi_{\delta}^{\mu' \sigma}\Phi_{\gamma}^{\rho \nu'}
+(\alpha\leftrightarrow\beta;\gamma\leftrightarrow\delta)\right]\right\}
m^{\dagger}_{\alpha}m^{\dagger}_{\beta}m_{\delta}m_{\gamma}\,.
\eea
Clearly, when the $\Phi$'s are the eigenstates of $H$, this will lead to an
expression that is equal to and the opposite sign of $V^{int}_{mm}$ in 
Eq.~(\ref{Vintra}). 

For the baryons, the exact same procedure is taken. The fourth order
$O$ operator is given by
\bea
O^{(4)}_{\alpha}=\frac{3}{8}\,\Delta_{\alpha\beta}\Delta_{\beta\gamma}
B_{\gamma}
-\frac{1}{8}\,B^{\dagger}_{\beta}[\Delta_{\alpha\gamma},\Delta_{\beta\delta}]
B_{\gamma}B_{\delta}
+\frac{1}{4}\,B^{\dagger}_{\beta}\{B_{\alpha},
[\Delta_{\beta\gamma},B^{\dagger}_{\delta}]\}B_{\gamma}B_{\delta}\,.
\label{bar4th}
\eea
The use of the Jacobi identity for fermionic operators $A,B,$ and $C$,
\beq
[A,\,\{B,C\}\,]+[C,\,\{A,B\}\,]+[B,\,\{C,A\}\,]=0,
\eeq
is useful for demonstrating the anticommutation relation of Eq.~(\ref{Oanti}).
As in the case of mesons, the cancelation of the intra-exchange
part of the effective baryon-baryon interaction is attained.

Next, the RGM is used with an exactly soluble model to demonstrate that the 
main effect of the orthogonality correction for the meson-meson effective 
interaction is the cancelation of the intra-exchange term. We consider the 
scattering of two mesons, where the quark and the antiquark have masses 
$m_q$, and use an harmonic potential for the microscopic interaction in 
Eq.~(\ref{mic}), namely, 
\beq
U(\bp) = - \int \frac{d\br}{(2\pi)^3} \, e^{-i\bp\bcdot\br}
\left(C+\frac{1}{2}k\br^2\right)
= \left(\frac{1}{2}k{\bnabla_{\bp}}^2-C\right)\delta^{(3)}(\bp) ,
\eeq
where $C$ is a constant which fixes the oscillator's ground state. For this
interaction, the Fock-space amplitude $\Phi$ is given by 
Eq.~(\ref{quark_mes_wf}), with  $b^2=\sqrt{3/2m_q k}$, and the total energy of 
a single meson is 
\beq
E(\bP)=\frac{\bP^2}{4m_q}+2m_q+\frac{3}{m_q b^2}+\frac{4C}{3}.
\eeq

The evaluation of normalization kernel and its square root can be done 
analytically. The results are,
\bea
&&N(\alpha\beta;\gamma\delta) =
\delta^{(3)}({\bP}_{\alpha}-{\bP}_{\gamma})
\delta^{(3)}({\bP}_{\beta}-{\bP}_{\delta})
-\frac{1}{6}\,
\caln_{E}({\bP}_{\alpha}{\bP}_{\beta};{\bP}_{\gamma}{\bP}_{\delta})\,,\\ 
&&N^{-\frac{1}{2}}(\alpha\beta;\gamma\delta) =
\delta_{\alpha\gamma}\delta_{\beta\delta}
\delta^{(3)}({\bP}_{\alpha}-{\bP}_{\gamma})
\delta^{(3)}({\bP}_{\beta}-{\bP}_{\delta})
+C_N\caln_{E}({\bP}_{\alpha}{\bP}_{\beta};{\bP}_{\gamma}{\bP}_{\delta}),
\label{knex}
\eea
where 
\beq
\caln_{E}(\bP_{\alpha}\bP_{\beta};\bP_{\gamma}\bP_{\delta})
=\delta^{(3)}(\bP_{\alpha}+\bP_{\beta}
-\bP_{\gamma}-\bP_{\delta})  
\left(\frac{b^2}{2\pi}\right)^\frac{3}{2}
e^{-\frac{b^2}{4}\Bigl[\bP^2_{\alpha}+\frac{\bP^2_{\gamma}}{2}+
\frac{\bP^2_{\delta}}{2}-\bP_{\alpha}\bcdot
(\bP_{\gamma}+\bP_{\delta})\Bigr]}.
\eeq
with
\beq
C_N=\frac{\omega}{2}\lim_{k\rightarrow \infty} 
\sum_{m=1}^{k}{\left(\frac{\omega}{2}\right)}^{m-1}
\left(\prod_{n=1}^{m}\frac{2n-1}{n}\right) ,
\label{cn}
\eeq
where $\omega=1/6$. In obtaining the closed form for the  square root 
$N^{-\frac{1}{2}}(\alpha\beta;\gamma\delta)$, we used the remarkable property
of the function $\caln_{E}$, that $(\caln_{E})^{k}=\caln_{E}$.

The partial sums $C(k)$,
\beq
C(k)= \sum_{m=1}^{k}{\left(\frac{\omega}{2}\right)}^{m-1}
\left(\prod_{n=1}^{m}\frac{2n-1}{n}\right),
\eeq
are plotted in Figure 2 below. The important fact to notice in this figure is
that the series is rapidly convergent, and that for $k \ge 2$, the values of 
the $C(k)$'s have almost reached their asymptotic value, 
$C(\infty) \sim 1.145$. This means that the bulk of the effect of the 
orthogonality corrections can be accounted for by retaining only the first 
term in Eq.~(\ref{expN-1/2}).

\begin{center}
\epsfxsize=10.0cm
\centerline{\epsfbox{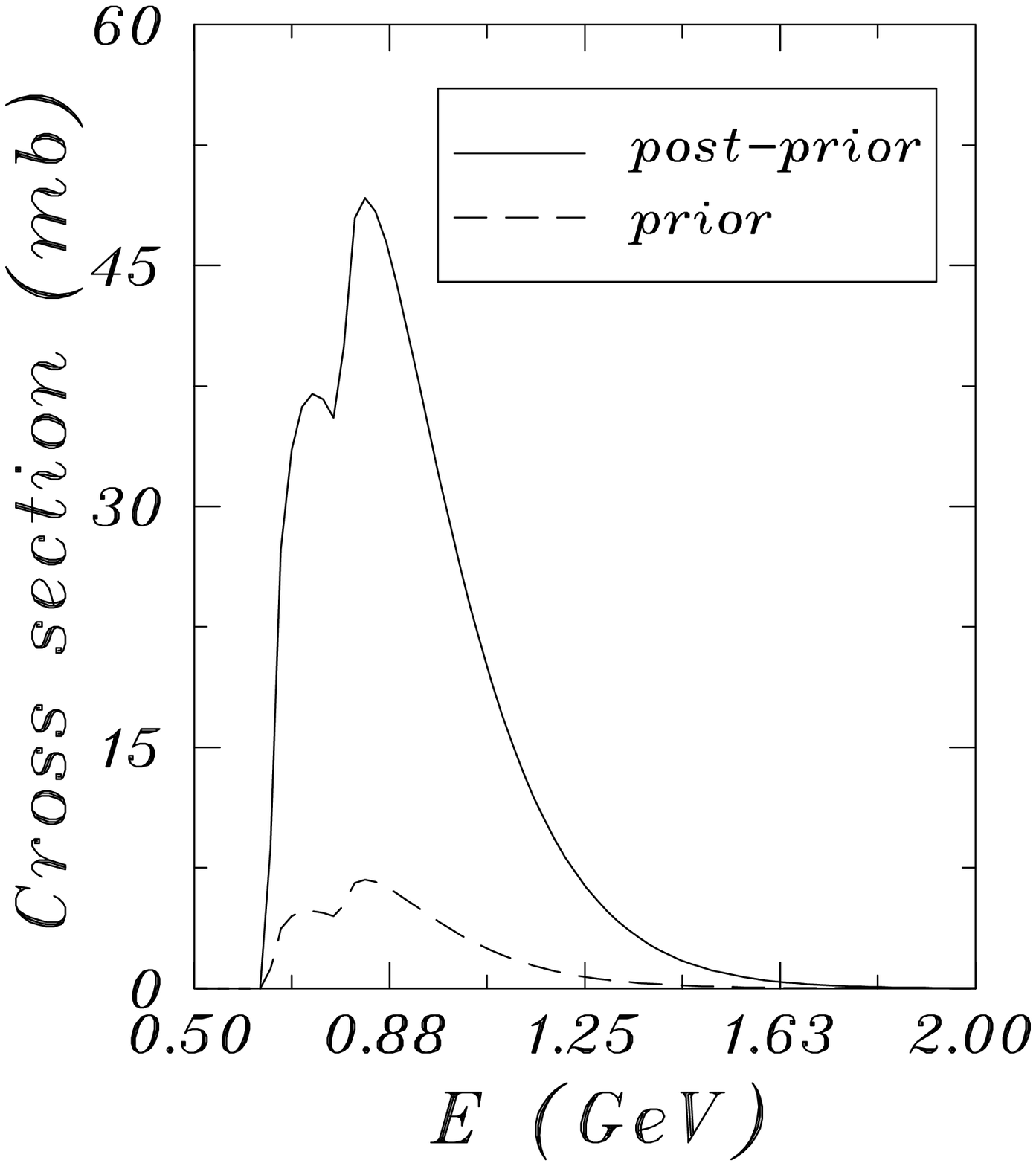}}
Figure 2. The partial sums $C(k)$ of Eq.~(\ref{cn}).
\end{center}

\vspace{0.5cm}
The full RGM equation can be separated into two equations, one describing the 
free motion of the c.m. of the mesons, and the other describing their relative 
motion. Writing ${\bar\Psi}(\bP_{\Lambda};\bp) = \psi(\bP_{\Lambda}) 
\varphi(\bp)$, where $\bP_{\Lambda}$ and $\bp$ are respectively the total c.m.
and relative momenta of the two mesons, the equation for 
$\varphi(\bp)$ can be written as
\beq
\int d\bp'\, \left[\frac{\bp^2}{M} \delta^{(3)}(\bp-\bp')+
\lambda \, {\cal V}(\bp,\bp')\right]\varphi(\bp')=E_{rel}\,\varphi(\bp) ,
\eeq
where $E_{rel}=E_{\Lambda}-E_{cm}$, and the ``potential" ${\cal V}(\bp,\bp')$,
which is the result from the renormalization of the kinetic and potential 
($V^{int}$ and $V^{exc}$) energies, is given by
\beq
{\cal V}(\bp,\bp')= \left(\frac{b^2}{2\pi}\right)^\frac{3}{2}\,
\left[\frac{b}{4 m_q}-\frac{1}{24m_q}(\bp^2+\bp'^2)\right]
e^{-\frac{1}{4}(\bp^2+\bp'^2)}\,,
\eeq
and $\lambda = C_K +C_V$, with
\beq
C_{K}=6 \,\frac{2{(1-\omega)}^{\frac{1}{2}}
+\omega-2}{{(1-\omega)}^{\frac{1}{2}}} \,,
\hspace{1.5cm}
C_{V}=\frac{1}{{(1-\omega)}^{\frac{1}{2}}}\,.
\eeq
Since $\omega=1/6$, $C_{V}+C_{K}=1.046$ is obtained. Had we used only the 
first two terms in the expansion of $N^{-1/2}$ in Eq.~(\ref{expN-1/2}), 
we would have obtained $\lambda=1$, instead of $\lambda=1.046$. 
The effect of the higher order terms is, therefore, less than $5\%$. 

This result stands for the case of an harmonic oscillator potential and 
gaussian Fock-space amplitudes $\Phi$, but it seems reasonable to expect that 
for other potentials and amplitudes the situation will not be extremely 
different from the present one. Of course, a check of the rate of
convergence of the expansion in Eq.~(\ref{expN-1/2}) is advisable when other 
than gaussian functions are used for the $\Phi$'s. 

For the case of baryons, when using a gaussian form for the Fock-space
amplitudes $\Psi$, the result is not different from the one for the
mesons as described above. The net result of the higher order terms in the 
expansion in Eq.~(\ref{expN-1/2}) is very small on the effective baryon-baryon
potential. However, contrary to the case of the mesons, the exchange kernel 
for baryons does not have the property $(\caln_{E})^{k}=\caln_{E}$. 
Nevertheless, there is an interesting approximation scheme that might be
useful for future calculations, as we shall shortly describe.

The normalization kernel for baryons is given by,
\bea
N(\alpha\beta;\gamma\delta ) &=& \delta_{\alpha \gamma}\,
\delta_{\beta \delta} - N_E(\alpha\beta;\gamma\delta) =
\delta_{\alpha \gamma}\,\delta_{\beta \delta}\,
- 9 \Psi _{\alpha}^{*\mu _1\mu _2\mu _3}
\Psi _{\beta}^{*\nu _1\nu _2\nu _3}
\Psi_{\gamma} ^{\nu _1\nu _2\mu _3}
\Psi _{\delta} ^{\mu _1\mu _2\nu _3} \nn\\
&\equiv& \delta_{\alpha \gamma}\,
\delta_{\beta \delta} - \,\omega\, 
\caln_{E}(\bp_\alpha \bp_\beta;\bp_\gamma\bp_\delta).
\label{aplic_1}
\eea
In the nucleon-nucleon case, $\omega$ is given by
\bea
\omega  \equiv
\frac{3}{4}\left[1+\frac{1}{9} \bm\tau^{1}_{N}\cdot\bm\tau^{2}_{N}
+ \frac{1}{9}\left(1+\frac{25}{9}\bm\tau^{1}_{N}\cdot\bm\tau^{2}_{N}
\right)\bm\sigma^{1}_{N}\cdot\bm\sigma^{2}_{N}\right].
\label{aplic_21}
\eea
The approximation consists in factorizing the spatial part and summing the 
resulting series as
\bea
N^{-\frac{1}{2}} &=&
\left[1- N_{E}\right] ^{-\frac{1}{2}}
=1+ \frac{1}{2}\,\omega \caln_{E} +
\frac{3}{8}\,\omega^{2} \caln_{E}^{2}+
\frac{15}{48}\,\omega^{3} \caln_{E}^{3}
+\ldots \label{exact} \nn\\
&\approx& 
1 + \caln_{E} \left[\frac{1}{2}\,\omega+\frac{3}{8}\,\omega^{2}+
\frac{15}{48}\,\omega^{3}+\ldots\right]
=1 +\vpi \,\caln_{E} ,
\label{approx}
\eea
where $\vpi= -1 + 1/\sqrt{1-\omega}$. When using a gaussian form for the 
amplitudes $\Psi$, two interesting facts were observed: first, there is a 
fast convergence of the exact and approximate series, and second, after the 
second term in the expansions, the approximate and exact results are 
practically indistinguishable from each other. Of course, as mentioned above, 
when other forms than a gaussian are used for the $\Psi$'s, the rate of
convergence of the expansion of the square root has to be checked.

\section{Extension to general Fock-space states}
\label{sec:gen}

In this section an extension of the Fock-Tani transformation to 
more general Fock-space states will be discussed. When dealing with quantum 
field theoretic quark models, the description of mesons and baryons might 
require the consideration of more general Fock states as those of a 
quark-antiquark pair and triplets of quarks. A meson state, for example, 
would be more likely to be of the form 
$\phi_1\,|q\aq> + \phi_2\,|q\aq(q\aq q\aq)> + \cdots$ and certainly mixing 
with states containing gluons, such as $|q\aq g>$, can be expected. Of course,
methods can always be devised to take into account as much of such a mixing as
possible into an effective Hamiltonian, and to avoid complicated Fock 
amplitudes. Nevertheless, one must be prepared to deal with more complicated 
Fock-space states as well. 

The unitary transformation for such states is constructed with the same
iterative procedure described in the previous sections, but the construction
of the generator $F$ requires modifications. In the present section, the
necessary modifications will be discussed through an example, and these will
be expanded upon in a future publication~\cite{BHGK} dedicated to the details 
of these derivations.

Consider a meson state of the form
\beq
M^{\dag}_{\alpha}=\Phi_{\alpha}^{\mu\nu}q_{\mu}^{\dag}\aq_{\nu}^{\dagger}+
\Psi_{\alpha}^{\mu\nu\sigma}q_{\mu}^{\dagger}\aq_{\nu}^{\dagger}
g_{\sigma}^{\dagger},
\label{exm2}
\eeq
where $g_{\sigma}^{\dagger}$ ($g_{\sigma}$) is a gluon creation (annihilation)
operator. The quark and antiquark operators satisfy the usual canonical
anticommutation relations of Eq.~(\ref{qcom}). For the sake of simplicity, we
assume the following canonical commutation  relations
\beq
[g_\sigma, g^{\dagger}_{\sigma'}]=\delta_{\sigma\sigma'},\hspace{1.0cm}
[g_\sigma, g_{\sigma'}]=[g^{\dagger}_\sigma, g^{\dagger}_{\sigma'}]=0.
\label{gcom}
\eeq
Using these and Eq.~(\ref{qcom}), the following expression is obtained for
the commutator of composite-meson operators
\beq
[M_{\alpha}, M_{\alpha'}^{\dag}]=\delta_{\alpha\alpha'} + C_{\alpha\alpha'},
\label{Mcomgen}
\eeq
where $C_{\alpha\alpha'}$ contains, in addition to the $\Delta_{\alpha\beta}$
of Eq.~(\ref{delta}), terms proportional to the amplitude 
$\Psi_{\alpha}^{\mu\nu\sigma}$ defined in Eq.~(\ref{exm2}). It is useful to 
decompose $C_{\alpha\alpha'}$ as
\beq
C_{\alpha\alpha'} = C_{\alpha\alpha'}^{0} + C_{\alpha\alpha'}^+ + 
C_{\alpha\alpha'}^-.
\eeq
$C_{\alpha\alpha'}^{0}$ contains the usual $\Delta_{\alpha\beta}$ and 
operators with at least one annihilation operator on the right,
\bea
&&C_{\alpha\alpha'}^{0}=-\Delta_{\alpha\beta}
+\Psi_{\alpha}^{*\mu \nu \sigma}\Psi_{\alpha'}^{\mu' \nu' \sigma}\,
q_{\mu'}^{\dag}\aq_{\nu'}^{\dag}\aq_{\nu}q_{\mu} 
+\Psi_{\alpha}^{*\mu \nu \sigma}\Psi_{\alpha'}^{\mu \nu \sigma'}\,
g_{\sigma'}^{\dag}g_{\sigma} 
- \Psi_{\alpha}^{*\mu \nu \sigma}\Psi_{\alpha'}^{\mu' \nu \sigma'}\,
q_{\mu'}^{\dag}g_{\sigma'}^{\dag}g_{\sigma}q_{\mu}\nn\\
&&\;\;\;-\Psi_{\alpha}^{*\mu \nu \sigma}\Psi_{\alpha'}^
{\mu' \nu \sigma}\,q_{\mu'}^{\dag}q_{\mu}
-\Psi_{\alpha}^{*\mu \nu \sigma}\Psi_{\alpha'}^{\mu \nu' \sigma'}\,
\aq_{\nu'}^{\dag}g_{\sigma'}^{\dag}g_{\sigma}\aq_{\nu}
-\Psi_{\alpha}^{*\mu \nu \sigma}\Psi_{\alpha'}^{\mu \nu' \sigma}\,
\aq_{\nu'}^{\dag}\aq_{\nu} .
\label{C0}
\eea
$C_{\alpha\alpha'}^+$ and $C_{\alpha'\alpha}^-$ contain only terms that
involve at least one $\Psi_{\alpha}^{\mu\nu\sigma}$,
\beq
C_{\alpha\alpha'}^+= \left[C_{\alpha'\alpha}^-\right]^{\dagger}
=\Phi_{\alpha}^{*\mu \nu}\Psi_{\alpha'}^{\mu\nu\sigma}\,
g_{\sigma}^{\dag}
-\Phi_{\alpha}^{*\mu \nu}\Psi_{\alpha'}^{\mu' \nu \sigma}\, 
g_{\sigma}^{\dag}q_{\mu'}^{\dag}q_{\mu}
-\Phi_{\alpha}^{*\mu \nu}\Psi_{\alpha'}^{\mu \nu' \sigma}\,g_{\sigma}^{\dag} 
\aq_{\nu'}^{\dag}\aq_{\nu}\,,
\label{Cpm}
\eeq

It is not difficult to prove that the operator $O_{\alpha}$ that satisfies the
commutation relations of Eq.~(\ref{comO}) up to third order in the Fock 
amplitudes is given by
\beq
O_{\alpha}=M_{\alpha}-\sum_{\alpha'}\left(C_{\alpha\alpha'}^++
\frac{1}{2} C_{\alpha\alpha'}^0\right)M_{\alpha'}+
\sum_{\alpha'\alpha''}M_{\alpha'}^{\dag}
\left[M_{\alpha},\left(C_{\alpha'\alpha''}^- + \frac{1}{2}
C_{\alpha'\alpha''}^0 \right)\right]M_{\alpha''} .
\eeq
The modification referred to above is that the components 
$C_{\alpha\alpha'}^{0}$, $C_{\alpha\alpha'}^+$ and  
$C_{\alpha\alpha'}^-$ of 
$C_{\alpha\alpha'}$ enter in a particular way into this expression. This is
because the commutator of $O_{\alpha}$
with $M^{\dag}_{\beta}$ must result in a Kronecker $\delta_{\alpha\beta}$, 
and the remaining operators of this commutator must annihilate the vacuum 
state, since $F\,M^{\dag}_{\alpha}|0>=m^{\dag}_{\alpha}|0>$ is needed. 
Since $C_{\alpha\alpha'}^+$ does not annihilate
the vacuum, the components of $C_{\alpha\alpha'}$ must be combined such that
they cancel out any operators which contain a creation operator on the right. 

To conclude, we mention that the same iterative procedure outlined for the 
above example can be followed~\cite{BHGK} for creation operators 
$A^{\dag}_{\alpha}$ involving any number of quark, antiquark and gluon 
(or other bosonic) creation operators in the form
\bea
A^{\dag}_{\alpha}&=&\sum_{n_q=0}^{\infty}\sum_{n_{\aq}=0}^{\infty}
\sum_{n_g=0}^{\infty}(n_q! n_{\aq}! n_g!)^{-1/2}
\Phi_{\alpha}^{\mu_1,\dots,\mu_{n_q},\nu_1,\dots,\nu_{n_{\aq}}, \sigma_1,
\dots,\sigma_{n_g}}\nn\\
&&\times\,q_{\mu_1}^{\dag}\,\cdots\, q_{\mu_{n_q}}^{\dag}\,\aq_{\nu_1}^{\dag}
\,\cdots\,\aq_{\nu_{n_{\aq}}}^{\dag}\,g_{\sigma_1}^{\dag}\,\cdots\,
g_{\sigma_{n_g}}^{\dag},
\label{creation} 
\eea
where $n_q$, $n_{\aq}$ and $n_g$ respectively represent the number of
quarks,  antiquarks and gluons. $\Phi_{\alpha}^{\mu_1,\dots,\mu_{n_q},
\nu_1,\dots, \nu_{n_{\aq}},\sigma_1,\dots,\sigma_{n_g}}$ is the Fock-space
amplitude, the index $\alpha$ represents the quantum numbers of the hadron and
$\mu$, $\nu$ e $\sigma$ those of the constituents, 
as usual. Such an extension is particularly useful for treating meson-baryon 
couplings, and the derivation of an one-boson-exchange picture of the 
nucleon-nucleon force. Work is in progress in this direction.

\section{Conclusions}
\label{sec:concl}

In this paper, we have extended the Fock-Tani representation to hadronic 
physics. The formalism was used for a general class of constituent quark 
models, independently of a specific form of the microscopic interaction. 
We derived the unitary transformation iteratively as a power series in the 
Fock-space hadron amplitude, obtained the transformed quark and antiquark 
operators, and derived effective Hamiltonians. The effective Hamiltonians
are hermitian and describe all possible interactions among the composite 
hadrons, and the interactions of the composite hadrons with their elementary 
constituents (quarks and antiquarks), consistent with the microscopic 
Hamiltonian. The transformed quark and 
antiquark operators are completely general, they depend only on the 
quark structure of the hadron states. Given a microscopic quark Hamiltonian, 
the effective Hamiltonians can be immediately derived. There is no 
restriction to relativistic or nonrelativistic kinematics. The method can be 
used with models where explicit gluon degrees of freedom (or other degrees of 
freedom such as Goldstone bosons) are present in the states and in the 
microscopic Hamiltonian, as well as with models of multicomponent Fock-space 
amplitudes.

The fact that in the Fock-Tani representation all field operators
satisfy canonical commutation relations allows the direct use of the
traditional field theoretic methods. For the baryon case, in particular, great 
opportunities for applications in many-baryon systems are envisioned.
Different methods have been employed in the literature to study various 
aspects of quark degrees of freedom in nuclei. Since the traditional picture 
of the nucleus is that of a system of hadrons, the explicit dynamics of the 
color degree of freedom must be limited to very short distances. This means 
that an approach using quark degrees of freedom should  minimally deviate 
from, as well as contain in some limit, the traditional one. In this sense, 
the effective Hamiltonian of the Fock-Tani representation has a well-defined
limit, since it explicitly describes the interactions among hadrons; 
quark-quark and quark-hadron interactions are treated separately as
``residual" interactions, since the effects of the bound states are explicitly
subtracted from the microscopic interaction. Herewith the properties of the
hadron-hadron interactions and nuclei at high densities and/or temperatures 
using quark degrees of freedom can be carried out in a systematic
and controllable fashion. 

\centerline{\bf Acknowledgments}
The authors would like to express their gratitude to Professor Marvin Girardeau
for his help in clarifying several issues of the Fock-Tani representation. The
authors acknowledge discussions with M. Betz, C. Maekawa and M.R. Robilotta. 
The present work was partially supported by 
the Alexander von Humboldt Foundation (Germany) and 
the Brazilian agencies CAPES, CNPq, FAPERGS and FAPESP.

\end{document}